\pdfoutput=1 
\documentclass[a4paper,twoside,11pt,fleqn]{article}
\usepackage{helvet}

\usepackage[latin1]{inputenc}
\usepackage[T1]{fontenc}
\usepackage{amsfonts}
\usepackage{amsmath}
\usepackage{amssymb,marvosym,units}
\usepackage{amsthm}
\usepackage[cmtip,arrow, matrix, curve]{xy}
\usepackage{pb-diagram,pb-xy}

\usepackage{graphicx}
\usepackage[svgnames,rgb]{xcolor}
\usepackage{epsfig}
\usepackage{tensor}

\usepackage{hyperref}
\usepackage{zref, xkeyval, ifpdf, ifthen, calc, marginnote, pdfcomment}

\usepackage{bbm}

\hypersetup{
    pagebackref=true,
    bookmarks=true,         
    unicode=false,          
    pdftoolbar=true,         
    pdfmenubar=true,       
    pdffitwindow=false,     
    pdfstartview={FitH},    
    pdftitle={Algebras of Quantum Variables for Loop Quantum Gravity},    
    pdfauthor={Diana Kaminski}
    pdfsubject={},   
    pdfkeywords={Loop Quantum Gravity and Cosmology}, 
    pdfnewwindow=true,      
    colorlinks=false,       
    linkcolor=black,          
    pdfborder= 0 0 0,
    pagebackref=true,
}

\title{Algebras of Quantum Variables for Loop Quantum Gravity\\[5pt]
\textbf{I. Overview}}
\author{Diana Kaminski\\[3pt]
kaminski@math.uni-paderborn.de\\ 
\small{Europe - Germany}}
\date{August 19, 2011}

\newcommand{\A}{\begin{large}\mathcal{A}\end{large}}
\newcommand{\Ab}{\begin{large}\bar{\mathcal{A}}\end{large}}

\newcommand{\Alg}{\begin{large}\mathfrak{A}\end{large}}

\newcommand{\Aut}{\begin{large}\mathfrak{Aut}\end{large}}

\newcommand{\CB}{\mathbb{C}}
\newcommand{\CD}{\mathcal{C}}

\newcommand{\DD}{\mathcal{D}}

\newcommand{\E}{\mathcal{E}}

\newcommand{\Goid}{\mathcal{G}}

\newcommand{\HS}{\mathcal{H}}

\newcommand{\KD}{\mathcal{K}}

\newcommand{\la}{\langle}
\newcommand{\LD}{\mathcal{L}}

\newcommand{\MD}{\mathcal{M}}
\newcommand{\MK}{\mathfrak{M}}
\newcommand{\MM}{\textbf{M}}

\newcommand{\ND}{\mathcal{N}}
\newcommand{\op}{\mathfrak{o}}

\newcommand{\PD}{\mathcal{P}}

\newcommand{\Ss}{\mathcal{S}}

\newcommand{\ra}{\rangle}

\newcommand{\FD}{\mathcal{F}}

\newcommand{\OD}{\mathcal{O}}

\newcommand{\QD}{\mathcal{Q}}
\newcommand{\R}{\mathbb{R}}

\newcommand{\WD}{\mathcal{W}}

\newcommand{\ZD}{\mathcal{Z}}

\newcommand{\ho}{\mathfrak{h}}

\newcommand{\go}{\mathfrak{g}}

\DeclareMathOperator{\dif}{d}

\DeclareMathOperator{\disc}{d}
\DeclareMathOperator{\Diff}{Diff}

\DeclareMathOperator{\LG}{LG}

\DeclareMathOperator{\phys}{phys}

\DeclareMathOperator{\Rep}{Rep}

\DeclareMathOperator{\tr}{tr}

\newcommand{\Gp}{{\Gamma^\prime}}

\newcommand{\Gpp}{\Gamma^{\prime\prime}}
\newcommand{\Gppp}{\Gamma^{\prime\prime\prime}}

\newcommand{\VC}{\overset{_\rightarrow}{C}}
\newcommand{\VN}{\overset{_\rightarrow}{N}}

\newcommand{\idf}{\mathbbm{1}}

\newcommand{\bra}{[}
\newcommand{\ket}{]}

\newcommand{\beq}{\begin{equation}\begin{aligned}}
\newcommand{\beqs}{\begin{equation*}\begin{aligned}}
\newcommand{\be}{\begin{flalign}}
\newcommand{\bes}{\begin{equation*}}
\newcommand{\eq}{\end{aligned}\end{equation}}
\newcommand{\eqs}{\end{aligned}\end{equation*}}
\newcommand{\ee}{\end{flalign}}
\newcommand{\ees}{\end{equation}}

\newcommand{\limN}{\lim_{N\rightarrow \infty}}

\newcommand{\limT}{\lim_{T\rightarrow\infty}}

\newcounter{exa}[section]

\newcounter{problem}[subsection]
 
\newcommand{\GGi}{\xymatrix{
  \Goid_1  \ar@<-2pt>[r] \ar@<2pt>[r] &  \Goid^0_1    \\
}}
\newcommand{\GGii}{\xymatrix{
  \Goid_2  \ar@<-2pt>[r] \ar@<2pt>[r] &  \Goid^0_2    \\
}}
\newcommand{\GGm}{\xymatrix{
  \Goid  \ar@<-1pt>[r]^{s} \ar@<1pt>[r]_{t} &  \Goid^0    \\
}}
\newcommand{\GGim}{\xymatrix{
  \Goid_1  \ar@<-1pt>[r]^{s_1} \ar@<1pt>[r]_{t_1} &  \Goid^0_1    \\
}}
\newcommand{\GGiim}{\xymatrix{
  \Goid_2  \ar@<-1pt>[r]^{s_2} \ar@<1pt>[r]_{t_2} &  \Goid^0_2    \\
}}

\newcommand{\PGm}{\xymatrix{
  \PD  \ar@<-1pt>[r]^{s} \ar@<1pt>[r]_{t} &  \Sigma    \\
}}

\newcommand{\fPGm}{\xymatrix{
  \PD_\Gamma  \ar@<-1pt>[r]^{s} \ar@<1pt>[r]_{t} &  V_\Gamma    \\
}}
\newcommand{\PGsm}{\xymatrix{
  \PD\Sigma \ar@<-1pt>[r]^{s_{\PD\Sigma}} \ar@<1pt>[r]_{t_{\PD\Sigma}} &  \Sigma   \\
}}
\newcommand{\fPGms}{\xymatrix{
  \PD^s_\Gamma  \ar@<-1pt>[r]^{s} \ar@<1pt>[r]_{t} &  V_\Gamma    \\
}}

\newcommand{\fPSGm}{\xymatrix{
  \PD_\Gamma\Sigma  \ar@<-1pt>[r]^{s} \ar@<1pt>[r]_{t} &  V_\Gamma    \\
}}
\newcommand{\fPSG}{\xymatrix{
  \PD_\Gamma\Sigma  \ar@<-2pt>[r] \ar@<2pt>[r] &  V_\Gamma    \\
}}
\newcommand{\fgHGm}{\xymatrix{
  H(\Gamma)  \ar@<-1pt>[r]^/0.3em/{\hat s_H} \ar@<1pt>[r]_/0.3em/{\hat t_H} &  V_\Gamma    \\
}}

\newcommand{\fHGm}{\xymatrix{
  H_\Gamma  \ar@<-1pt>[r]^/0.3em/{\hat s_H} \ar@<1pt>[r]_/0.3em/{\hat t_H} &  V_\Gamma    \\
}}

\newcommand{\fGGm}{\xymatrix{
  \G^G_\Gamma  \ar@<-1pt>[r]^/0.3em/{s_P} \ar@<1pt>[r]_/0.3em/{t_P} &  V_\Gamma    \\
}}
\newcommand{\fGHm}{\xymatrix{
  \G^H_\Gamma  \ar@<-1pt>[r]^/0.3em/{s_P} \ar@<1pt>[r]_/0.3em/{t_P} &  V_\Gamma    \\
}}

\newcounter{count}
\newcommand{\citetableA}{\cite[table 1]{Kaminski1} or \cite[table 11.1]{KaminskiPHD}}
\newcommand{\citetableB}{\cite[table 11.2]{KaminskiPHD}}
\newcommand{\citetableC}{\cite[table 11.3]{KaminskiPHD}}
\newcommand{\citetableD}{\cite[table 11.4]{KaminskiPHD} or \cite[Section 6]{Kaminski2}}

\newcommand{\citetableF}{\cite[table 11.6]{KaminskiPHD}}
\newcommand{\refpropmultilpiercrossprod}{Theorem 7.2.9 }
\newcommand{\refsecAPLQC}{Chapter 4}
\newcommand{\refsecalgperiod}{Section 8.1}
\newcommand{\reftheoweylnokmsstates}{Theorem 6.5.8}
\newcommand{\reftheouniquweylalg}{Theorem 6.4.6}
\newcommand{\refsubsecholfluxdiffcrossalg}{Section 7.3}
\newcommand{\reftablecompalg}{table 11.5}

\begin{document}
\maketitle
\begin{abstract}
\noindent
The operator algebraic framework plays an important role in mathematical physics. Many different operator algebras exist for example for a theory of quantum mechanics. In Loop Quantum Gravity only two algebras have been introduced until now. In the project about \textit{Algebras of Quantum Variables (AQV) for LQG} the known holonomy-flux $^*$-algebra and the Weyl $C^*$-algebra will be modified and a set of new algebras will be proposed and studied. The idea of the construction of these algebras is to establish a finite set of operators, which generates (in the sense of Woronowicz, Schm\"udgen and Inoue) the different $O^*$- or $C^*$-algebras of quantum gravity and to use inductive limits of these algebras. In the Loop Quantum Gravity approach usually the basic classical variables are connections and fluxes. Studying the three constraints appearing in the canonical quantisation of classical general relativity in the ADM-formalism some other variables like curvature appear. Consequently the main difficulty of a quantisation of gravity is to find a suitable replacement of the set of elementary classical variables and constraints. The algebra of quantum gravity is supposed to be generated by a set of the operators associated to holonomies, fluxes and in some cases even the curvature. There are two reasonable choices for this algebra: The set of constraints of Quantum Gravity are contained in or at least the constraints are affilliated with this algebra. Secondly, the algebra of quantum variables is said to be physical if it contains complete observables. 
In the project of \textit{Algebras of Quantum Variables for LQG} different algebras will be studied with respect to the property of being a physical algebra. Furthermore the existence of KMS-states on these algebras will be argued.\\

\noindent
Summarising this article will give an overview about the following objects 
\begin{itemize}
 \item the two known algebras for Quantum Gravity in the Loop Quantum Gravity approach:\\[3pt]
the holonomy-flux $^*$-algebra \cite{LOST06} and the Weyl $C^*$-algebra \cite{Fleischhack06}, 
 \item modifications of these algebras and new algebras for Loop Quantum Gravity,
 \item states and representations of the algebras and 
 \item a concept of quantum constraints and KMS-Theory in Loop Quantum Gravity.
\end{itemize}
\end{abstract}

\newpage
\thispagestyle{plain}
\pdfbookmark[0]{\contentsname}{toc}
\tableofcontents

\section{Introduction}

During the last years two quantum algebras for a gravitational quantum theory in the concept of Loop Quantum Gravity have been developed. The holonomy-flux $^*$-algebra has been introduced by the project group of Lewandowski, Okolow, Sahlmann and Thiemann \cite{LOST06} and the Weyl $C^*$-algebra has been presented by Fleischhack \cite{Fleischhack06}. Both algebras are generated by the quantised canonical variables of gravity, which are given by the holonomies and the fluxes. The fundamental aspect of  the holonomy-flux $^*$-algebra and the Weyl $C^*$-algebra is given by the uniqueness of the representation of these algebras with respect to diffeomorphism invariance and the unitary (weakly) continuous representation of the fluxes on some Hilbert space. In the project of \textit{Algebras of Quantum Variables in LQG (AQV)} the following questions are studied:
\begin{itemize}
 \item Which algebras can be generated by operators, which are derived from holonomies along paths and fluxes associated to surfaces, and relations among these operators? Moreover, which $^*$- or $C^*$-algebras can be constructed?
 \item Which are the basic classical variables, which need to be quantised, for a theory of quantum gravity with constraints?
 \item If holonomies along paths, fluxes, diffeomorphisms are quantised, then which algebras are generated by these quantum operators?
  \item If additional the classical curvature is concerned, which algebra is generated by the quantum operators and the quantum curvature? 
 \item If several algebras of quantum gravity are constructed, which algebra is the preferred one? 
 \end{itemize}

This article will give an overview about the project about \textit{Algebras of Quantum Variables in Loop Quantum Gravity}. The structure of the article is the following. The first chapter is about the quantisation procedure of classical gravity in the framework of Loop Quantum Gravity. A quantisation of a classical theory provides a set of quantum operators, which generate for example $^*$-, $O^*$-, $C^*$- or von Neumann algebras. A first motivation for a construction of different algebras is presented in the context of the Quantum Mechanics in subsection \ref{subsec QM}. The underlying mathematical theory is studied very briefly in section \ref{subsec maththeory}. An overview about the basic quantised variables of the theory are given in section \ref{subsec algLQG}. Then by using the mathematical framework the following different algebras for Loop Quantum Gravity will be considered in several articles:
\begin{enumerate}
 \item\label{unit3} a new formulation of the Weyl $C^*$-algebra for Loop Quantum Gravity\\ \cite{Kaminski1,KaminskiPHD}\\ 
 a new algebra - the holonomy-flux von Neumann algebra \cite{Kaminski4,KaminskiPHD}\\
 \item\label{unit4} new algebras - the flux group, flux transformation group, holonomy and heat kernel holonomy $C^*$-algebra\\ 
 a new algebra - the holonomy-flux cross-product $C^*$-algebra\\ \cite{Kaminski2,KaminskiPHD}
 \item\label{unit5} new analytic holonomy $^*$-algebras \cite{KaminskiPHD},\\
a new algebra - the holonomy-flux cross-product $^*$-algebra  \cite{Kaminski3,KaminskiPHD}, which is a comparable with the holonomy-flux $^*$-algebra \cite{LOST06},\\ a new algebra - the localised holonomy-flux cross-product $^*$-algebra \cite{Kaminski4,KaminskiPHD} and\\
other new holonomy-flux $^*$-algebras
 \cite{KaminskiPHD}\\
 \item\label{unit6} a new algebra - the holonomy groupoid $C^*$-algebra for a gauge theory\\ \cite{Kaminski5,KaminskiPHD}
\end{enumerate}
A short summary over the basic constructions and results in comparison to known algebras in LQG is presented in section \ref{subsec algLQG}. Moreover, an outline of the implementation of quantum constraints is illustrated in the framework of Loop Quantum Gravity and a motivation for the study of KMS-states is given in section \ref{sec Iquant}. Furthermore, the analysis will indicate that, a study of different quantum algebras is necessary for the application of quantum constraints. These concepts allow to identify a set of conditions for an exceptional algebra, which is called a physical algebra for a theory of quantum gravity in the framework of LQG.

\subsection{Quantisation procedures of a classical system}

In mathematical quantum physics the following quantisation procedures play a fundamental role:
\begin{enumerate}
 \item the Canonical Quantisation Procedure based on Hilbert space methods introduced by Dirac (1948/49)
 \item the Algebraic Quantisation of observables related to $^*$-, $C^*$-, von Neumann or generally Jordan algebras and 
 \item Path Integral Methods.
\end{enumerate}

In general, a classical Hamiltonian dynamical system with $n$ degrees of freedom is a $2n$-dimensional manifold $P$, 
which is called the phase space. Often the phase space is encoded in a cotangent bundle over a classical configuration space. The canonical variables are given by a set of $2n$ functions $(x_i,p_i)$ on the phase space for $i=1,...,n$. The Poisson bracket for functions of the space $C^\infty(P)$ of real-valued, infinitely differentiable functions defines a Poisson algebra. 

Dirac or Canonical Quantisation Procedure is given by the assignment of functions in $C^\infty(P)$ with symmetric operators on a Hilbert space $\HS$. In particular, the derivations of the associative Poisson algebra are replaced by a symmetric operators on a dense domain and all quantum operators (for example the constraints) are defined on a common invariant dense subspace of $\HS$. In some examples one can construct a quantisation map $\QD$ from a suitable subspace of the Poisson algebra to a suitable Lie algebra of symmetric operators on a Hilbert space on a common invariant dense domain. In the LQG approach to quantum gravity the classical algebra of position and momentum variables and constraints form a difficult algebra. Hence, the algebra of suitable symmetric operators on a Hilbert space is not easy to construct.

Another procedure is the operator algebraic quantisation of classical canonical variables. In this framework the quantum operators in a suitable algebra encodes the classical variables without referring to a Hilbert space. 
The new algebraic formulation allows to study a huge amount of different algebras. For a first overview about the ideas for a construction of different algebras of quantum variables simple physical examples are studied in the next section. 

\subsection{Algebras in Quantum Mechanics}\label{subsec QM}

First of all, quantum algebras can be generated by bounded or unbounded operators. Quantised configuration and momentum variables, which are bounded operators, generate $C^*$-algebras, whereas quantum operators, which are unbounded, form $O^*$-algebras. Later an extended theory of $C^ *$-algebras generated by unbounded operators are presented. In this section unbounded operators form $O^*$-algebras, which are certain $^*$-algebras defined on a dense domain. The concept of $O^*$-algebras can be found in \cite{Inoue}. Now, the diversity of quantum algebras in Quantum Mechanics is studied very briefly.

\paragraph{$O^*$-algebras of Quantum Mechanics\\}\hspace{10pt}

In Quantum mechanics the classical variables of the theory are position and momentum variables. The quantisation maps replace the position operator by a single projection-valued measure (PVM)\footnote{In other words there is a map $E\mapsto P_E$ from a Borel subset $E\subset \R^3$ to the projections on $L^2(\R^3)$ that satisfies $P_\varnothing =0$, $P_{\R^3}=1$, $P_EP_F=P_FP_E=P_{E\cap F}$ for all measurable $E,F\subset \R^3$, and $P_{\cup E_i}=\sum P_{E_i}$ for all countable collections of mutually disjoint $E_i\subset\R^3$} $P_E$ on $\R^3$ with values in the separable Hilbert space $L^2(\R^3)$ where $E\subset\R^3$. Then the classical variable $x$ is replaced by $P_E(x)$.  For a measurable function $f : \R^3 \rightarrow \R$ the quantised operator of a classical function $f$ is given by 
\[ \QD(f)\psi : = \int_{\R^3}\dif P_E(x) f(x)\psi\]
This operator acts as a multiplication operator on the Hilbert space and is self-adjoint on the domain \[D(f):=\left\{\psi\in L^2(\R^3):\int_{\R^3}\dif \langle\psi, P_E(x)\psi\rangle\vert f(x)\vert^2<\infty\right\}\] The quantum position operator is given by
\[ \QD(x^j)\psi : = \int_{\R^3}\dif P_E(x) x^j\psi\] 
and is defined on the Schwartz space $\Ss(\R^3)$ of all rapidly decreasing smooth functions on $\R^3$.

On the other hand, there exists a strongly continuous unitary group representation $V$ of the group $\R^3$ on the Hilbert space $L^2(\R^3)$, which implements translations on $\R^3$ by $ V(y)\psi(x) = \psi(x-y)$. Then the quantised momentum operators are represented on the Hilbert space by the assignment
\[\QD(p_i)\psi := i\hbar \lim_{x_i\rightarrow 0}x_i^{-1}\left(V(x_i)-1\right)\psi,\]
where $V(x_i):=V(0,...,0,x_i,0,...,0)$. The operators  $\QD(p_i)$ are self-adjoint on the set of all $\psi$ for which the limit exists (Stone theorem). The assignment of the quantum momentum operator $\QD(p_i)$ is also denoted by the operator $\dif V(x_i)$, which is called the infinitesimal representation $\dif V$ of the group $\R$ on the Hilbert space $L^2(\R^3)$, and defines the partial derivation $\partial_{x_i}$. 

The representation of the quantum operators $\QD(x^i):=x^i$ and $\QD(p_i):=i\hbar \partial_{x^i}$ on the Hilbert space $L^2(\R^3)$ is called the Schr\"odinger representation of quantum mechanics. The canonical commutation relations of the quantum operators is encoded in the relation $[\QD(p_i),\QD(x^j)]=-i\hbar \delta^j_i$. The bounded operator $\idf$ and the unbounded operators $x^i$ and $\partial_{x^j}$ for all $1\leq i,j\leq 3$ satisfying these relations generate an associative Lie $^*$-algebra or, in general, a closed operator $O^*$-algebra. The elements of this algebra are of the form \[\sum_{1\leq k\leq m}\sum_{1\leq l\leq n}\lambda_{kl}x^k\Big(\frac{\dif}{\dif x}\Big)^l\text{ for }\lambda_{kl}\in\CB\] This $O^*$-algebra is called the Heisenberg algebra $\mathfrak{O}_{\text{Heis}}$ of Quantum Mechanics. 

Furthermore, there exists another $^*$-algebra. The canonical commutator relations of the quantum operators $\QD(p_i):=i\hbar \partial_{x^i}$ and $\QD(f):=f$ for every function $f\in C^\infty(\R^3)$ are $\bra \QD(p_i),\QD(f)\ket=i\hbar \QD(X f)$, where the canonical vector field $X$ on $\R^3$ is defined by \[(Xf)(y):=\frac{\dif}{\dif t}\Big\vert_{t=0}f(\exp(-tx^i)y)\]
Then the algebra generated by $\QD(f)$ for every function $f\in C^\infty(\R^3)$ and $\QD(p_i)$ for $1\leq i\leq 3$, which satisfy the canonical commutator relations, is an $O^*$-algebra, too. The elements are of the form \[\sum_{1\leq k\leq m}\sum_{1\leq l\leq n}f_k(x)\Big(\frac{\dif}{\dif x}\Big)^l\text{ for every }f_k\in C^\infty(\R^3)\]

Hence to summarise two different $O^*$-algebras are illustrated by using only different functions of the configuration variables in Quantum Mechanics. In LQG approach the ideas will be used to define different $^*$-algebras.

\paragraph{$C^*$-algebras in Quantum Mechanics\\}\hspace{10pt}

$C^*$-algebras are isomorphic to norm-closed $^*$-subalgebras of the $C^*$-algebra of bounded operators on some Hilbert space. Consequently, bounded operaors on a Hilbert space can be used to define $C^*$-algebras.

Now, the quantisation map of the position operator $\QD(x)$ is  given by the unitary operator $u_x:=\exp(ix)$ and the quantum momentum $\QD(p)$ is equal to the unitary $v_p=\exp(-ip)$. Then the canonical commutation relation for the abelian locally compact group $\R^3$ changes to
\beq\label{eq cancomrelQM} v_pu_xv^*_p=\exp(i\langle x,p\rangle)u_x\eq   where $v^*_p=v_{-p}$ and $\langle.,.\rangle$ denotes the euclidean inner product on $\R^3$.  The $^*$-algebra generated by the Weyl elements $W(x,p):=\exp(i(x-p))$ and $W^*(x,p):=\exp(-i(x-p))$ satisfying the commutator relations can be completed in a $C^*$-norm. This completion is called the Weyl $C^*$-algebra of Quantum Mechanics. 

The quantum algebra generated by the configuration variables only is given by the continuous functions $C_0(\R^3)$ vanishing at infinity with pointwise multiplication and supremum norm. Notice that $C_0(\R^3)$ is isomorphic to the group algebra $C^*(\R^3)$. 

There is a $\R^3$-covariant representation $(\HS,V,\pi)$ of the $C^*$-algebra $C_0(\R^3)$ such that there exists an action $(L_pf)(y):=f(y-p)$ of $\R^3$ on $C_0(\R^3)$, a continuous unitary representation $V$ of $\R^3$ on the Hilbert space $\HS$, and a non-degenerate representation $\pi$ of $C_0(\R^3)$ on $\HS$ satisfying
\beq\label{eq cancomrelzwei} V(p)\pi(f)V(p)^{*}=\pi(L_pf)\text{ for all }p\in \R^3,f\in C_0(\R^3)\eq Consequently, another Weyl $C^*$-algebra can be constructed by the Weyl elements $v_p$ and a the $C^*$-algebra $C_0(\R^3)$, which satisfy the canonical commutator relation \eqref{eq cancomrelzwei}. 

There is another idea of a quantisation map. Consider the quantum position operator $\QD(f):=f$ for a function $f$ in $C_0(\R^3)$ depending on $x$ and the quantum momentum $\QD(m)$ for functions $m$ in $C(\R^3)$ depending on the momentum. In particular, the quantum momentum is defined by $\QD(m):=\int_{\R^3}\dif\mu(p)m(p)V(p)$, where $\mu$ is the Lebesque measure on $\R^3$. The quantum momentum can be generalised further. For example, let $f$ be an element in $L^1(\R^3,C_0(\R^3))$, which is a certain function from $\R^3$ to an algebra element of $C_0(\R^3)$. Then other $C^*$-algebras for Quantum Mechanics are available. The reduced cross-product $C^*$-algebra is the completion of the Banach $^*$-algebra $L^1(\R^3,C_0(\R^3))$ in the $L^2(\R^3)$-norm. The representation $\pi_I$ of the Banach $^*$-algebra $L^1(\R^3,C_0(\R^3))$ on the Hilbert space $L^2(\R^3)$ is given by
\beqs \pi_I(f)\psi=\int_{\R^3} \dif\mu(p) f(p) V(p)\psi\eqs whenever $f\in L^1(\R^3,C_0(\R^3))$ and $\psi\in L^2(\R^3,\dif\mu)$. The representation $\pi_I$ is called the integrated or generalised momentum representation. The general cross-product $C^*$-algebra is completion of the Banach $^*$-algebra $L^1(\R^3,C_0(\R^3))$ with respect to the universal norm. 

The unbounded operator $\dif V(x)$, which is defined as the infinitesimal representation $\dif V$ of the group $\R^3$,  is not contained in the cross-product $C^*$-algebra. These operators are affiliated operators with this $C^*$-algebra. In this context, for example Woronowicz \cite{Woro91,WoroNap,Woronowicz95} and Schm\"udgen \cite{Schmuedgen90,Schmuedgenweb} speak about the $C^*$-algebra, which is generated by the bounded quantum position operators $\QD(f)$ for every $f\in C_0(\R^3)$ and the unbounded quantum momentum operators $\QD(p):=i\hbar\dif V(x)$ for every $x\in\R^3$.

In particular, this cross-product $C^*$-algebra is a certain transformation group $C^*$-algebra, and is denoted by $C^*(\R^3,\R^3)$. Furthermore, this $C^*$-algebra is Morita equivalent to the $C^*$-algebra $\KD(L^2(\R))$ of compact operators. The representation theory of the $C^*$-algebra $\KD(L^2(\R))$ is rather simple, since there is only one irreducible representation up to unitary equivalence. Morita equivalence of $C^*$-algebras implies that the representation theories of both $C^*$-algebras are the same. Hence, all irreducible representations of the cross-product $C^*$-algebra are unitarily equivalent to $\pi_I$. This result generalises the famous Stone - von Neumann theorem about the uniqueness of the irreducible Schr\"odinger representation of the Weyl $C^*$-algebra of Quantum Mechanics.

The elements of the Weyl algebra satisfy the canonical commutator relation. Then the exponentiated euclidean inner product $\la.,.\ra$ in $\R^3$, which is denoted by $\sigma$, can be used for the definition of a new $C^*$-algebra, which is distinguished from the Weyl $C^*$-algebra of Quantum Mechanics. This algebra is called the twisted transformation group $C^*$-algebra $C^*_\sigma(\R^3,\R^3)$. The construction is similar to the transformation group $C^*$-algebra $C^*(\R^3,\R^3)$. 

Summarising, the following $C^*$-algebras have been presented:
\begin{itemize}
\item two Weyl $C^*$-algebras,
\item quantum $C^*$-algebra of position operators, which is isomorphic to the group $C^*$-algebra,
\item the cross-product $C^*$-algebra, which is also called the transformation group $C^*$-algebra of Quantum Mechanics, and
\item the twisted transformation group $C^*$-algebra.
\end{itemize}

The different quantum algebras derived from the concepts presented above will be used to define the Weyl $C^*$-algebra for surfaces, the analytic holonomy $C^*$-algebra and the holonomy-flux cross-product $C^*$-algebra in the Loop Quantum Gravity approach. 
 
\subsection{General mathematical concepts for the construction of $C^*$-algebras}\label{subsec maththeory}

In this section a general mathematical framework is presented, which generalises the examples for Quantum Mechanics. In particular some problematic aspects are collected.

\subsubsection{Pontryagin duality, quantum groups and cross-product algebras}\label{subsec Pontryaginduality}
\paragraph*{$C^*$-algebras of quantum configuration or momentum variables\\}\hspace{10pt}

It is well-known that for every locally compact abelian group $G$ the Pontryagin duality\footnotetext{Pontryagin duality states that there is a mapping $U:G\rightarrow \widehat{\hat G}$ defined by $U(\hat g)(g)=\hat g(g)$ for all $\hat g\in\hat G$ and $g\in G$, is a group isomorphism and a homeomorphism.} holds. Furthermore, in \cite{Blackadar,HewittRoss} the following arguments are given to conclude that there is an identification of unitary representations of an abelian locally compact group $G$ with representations of the algebra $C_0(\hat G)$ of functions on the Pontryagin dual $\hat{G}$ vanishing at infinity. As a set, $\hat G$ is the set of all continuous group characters on $G$ taking values in the unit circle. The group multiplication of $\hat G$ is the pointwise multiplication of characters. The topology of $\hat G$ is the compact-open topology, in which a net of elements in $\hat G$ converges to an element in $\hat G$ if the net converges uniformly to the element $\hat g$ on each compact subset of $G$. Clearly, $\hat G$ equipped with this structure is a commutative locally compact group. 

Let $C^*_r(G)$ be the reduced group $C^*$-algebra, which is generated by matrix elements of the fundamental representation of a abelian locally compact group $G$. Therefore, via the Gel'fand-Na\u{\i}mark theorem there is a $^*$-isomorphism (the Fourier-Plancherel Transform) between $C_0(\hat G)$ and $C^*_r(G)$. Define the following function $\hat f\in C_0(\hat G)$ by
\beq\label{Fourierablcag} \hat f(\hat g):=N\int_G \dif \mu_H(g) f(g)\la g,\hat g\ra \quad\forall f\in C^*_r(G)
\eq where $U_{\hat g}(g):=\la g,\hat g\ra$ defines a finite-dimensional unitary continuous representation of $G$ and $N$ is a suitable constant. Then there is an isomorphism $\FD: C^*_r(G)\rightarrow C_0(\hat G)$, which is defined by $f\mapsto \FD(f)=\hat f$. Note that for $G=\R$ the Fourier transform reads
\beqs \hat f(x):=N\int_\R \dif \mu(p) f(p)\exp(i\la p,x\ra)\quad\forall f\in C^*_r(\R),\hat f\in C_0(\R)
\eqs The pointwise product in $C_0(\hat G)$ is given by
\beqs \hat f(\hat g)\cdot  \hat k(\hat g)
=M\int_G \dif \mu_H(g) (f\ast k)(g)U_{\hat g}(g) \quad\forall f\in C^*_r(G)
\eqs  For general non-abelian locally compact groups there is in general no isomorphism between the two $C^*$-algebras.

Consequently, one has to argue which algebra is more fundamental or which algebra encodes the information about the group $G$. This problem has been studied by Woronowicz in \cite{Woro91}. First assume that $G$ is locally compact and abelian. Then Woronowicz considers two quantum groups, which are defined by the pairs $(C^*_r(G),\hat\bigtriangleup_r)$ and $(C_0(\hat G),\bigtriangleup)$, where $\hat\bigtriangleup_r$ and $\hat\bigtriangleup$ are comultiplications. A generalisation of \eqref{Fourierablcag} is given by the so called generalised Fourier transform
\beq\label{eq genFourrepres} \pi_I(f):= \int_G \dif \mu_H(g) f(g)U(g)
\eq whenever $U$ is a continuous unitary representation of $G$ on a Hilbert space $\HS$. The map $\pi_I$ is a representation of $C_r(G)$ on the Hilbert space $\HS$. Furthermore, the map $\pi_I$ is compatible with these quantum group structure and, therefore, $(C^*_r(G),\hat\bigtriangleup_r)$ and $(C_0(\hat G),\bigtriangleup)$ are isomorphic as quantum groups. For non-abelian locally compact groups the integrated representation of $C^*_r(G)$ can be defined by \eqref{eq genFourrepres}, too. But in this case, the $C^*$-algebra $C^*_r(G)$ is non-commutative and, hence, $(C^*_r(G),\hat\bigtriangleup_r)$ is not isomorphic (as quantum groups) to a quantum group $(C_0(H),\bigtriangleup)$ where $H$ is a locally compact group. The appropriate dual group $H$ of $G$ is not constructable. Woronowicz in \cite{Woro82} have argued that the non-commutative reduced group algebra $C^*_r(G)$ encodes as a quantum group all information about $G$. 
Summarising, for non-commutative locally compact groups either the group $C^*$-algebra $C^*(G)$ or the $C^*$-algebra $C_0(G)$ can be analysed. Unfortunately, there need not exists an isomorphism between the two algebras. 

Notice that if $G$ is a connected Lie group, the basis $T_1,..,T_N$ of the Lie algebra $\go$ of $G$ are skewadjoint unbounded operators, which are affiliated with the group $C^*$-algebra $C^*(G)$. Moreover, Woronowicz and Napi\'{o}rkowski have shown in \cite{WoroNap} that, the group $C^*$-algebra $C^*(G)$ can be generated by these unbounded elements (in the sense of Woronowicz or Schm\"udgen). The arguments can be generalised such that it can be used for example in the framework of the holonomy-flux cross-product $C^*$-algebra. In this project \textit{AQV} the classical flux variables are quantised either as $G$-valued quantum flux operators or Lie algebra-valued quantum flux operators. The latter is defined for a structure group $G$ being a compact Lie group and $\go$ denotes the associated Lie algebra. In this context, the Lie-algebra-valued quantum flux operators are affiliated operators with the holonomy-flux cross-product $C^*$-algebra. This algebra is constructed from the quantum configuration and quantum momentum operators, which are given by the holonomies along paths and the $G$-valued quantum flux operators associated to surfaces.

\paragraph*{The $C^*$-algebras of quantum configuration and quantum momentum variables\\}\hspace{10pt}

A general Weyl $C^*$-algebra is generated by Weyl elements satisfying some canonical commutator relations. Notice that there is no general definition of a Weyl algebra, since, the Weyl elements can be defined in many different ways. In the following the focus lies on abelian locally compact groups. The Weyl algebra $\WD$ is generated by Weyl elements, which can be constructed by an unitary continuous representation $\pi$ of $G$ and an unitary continuous representation $\Pi$ of $\hat{G}$ on a Hilbert space $\HS$ for which the canonical commutator relations 
\beq\label{eq cancomlcagroup}\pi(g)\Pi (\hat{g})\pi(g)^*=\langle g,\hat{g}\rangle \Pi(\hat{g})\eq is satisfied.   

In the case of an arbitrary non-commutative locally compact group $G$, the concept of a $G$-covariant representation $(\HS,U,\pi)$ corresponding to a $C^*$-dynamical system is useful. The $C^*$-dynamical system is a triple $(G,C_0(X),\alpha)$, which is given by a point-norm continuous automorphic action $\alpha$ defined by $(\alpha_xf)(y)=f(x^{-1}y)$ of $G$ on $C_0(X)$. The Weyl algebra is constructed from Weyl elements, which are defined by the unitary continuous representations of $G$ on a Hilbert space $\HS$. Each unitary representation $U$ and a representation of the $C^*$-algebra $C_0(X)$ define a covariant pair $(U,\Phi_M)$ such that
\beqs U(g)\Phi_M(f)U^*(g)=\Phi_M(\alpha(g)(f))
\eqs Clearly, if an abelian locally compact group is considered this Weyl algebra is equivalent to the Weyl algebra $\WD$ defined in the previous paragraph. 

For simplicity assume that the set $X$ is equal to $G$. Then a new $C^*$-algebra can be constructed from the $C^*$-dynamical system $(G,C_0(G),\alpha)$. This $C^*$-algebra is called the transformation group $C^*$-algebra $C^*(G,G)$. Furthermore, the $C^*$-algebra $C^*(G,G)$ is isomorphic to the algebra $\KD(L^2(G))$ of compact operators on $L^2(G)$. Rieffel generalises this result. This result is called the generalised Stone - von Neumann theorem and states that these $C^*$-algebras are Morita equivalent. Consequently, there is a theorem available, which ensures a uniqueness result for irreducible representations of the transformation group $C^*$-algebra.

Assume that $G$ is non-commutative locally compact group.
Then in this project \textit{AQV} the structures presented above will be used for the definition of the Weyl algebra associated to surfaces in the context of Loop Quantum Gravity. The set of $G$-valued quantum flux operators associated to a certain surface set and a graph forms a group. This group is called the \textit{flux group associated to a surface set and a graph}. Elements of this group are called \textit{$G$-valued quantum flux operators associated to surface}. Clearly, for different sets of surfaces many different flux groups exists. A restricted Weyl algebra is obtained for a certain graph and a certain set of suitable surfaces. Then the restricted configuration space $\Ab_\Gamma$ is derived from a set of holonomies along paths in a graph $\Gamma$ and the restricted momentum space is given by the flux group associated to the surface set and the graph $\Gamma$. The Weyl $C^*$-algebra associated to a surface set and a graph will be constructed from a set of holonomies along paths in a graph $\Gamma$, the $G$-valued quantum flux operators associated to surfaces in a surface set and the graph, and automorphic actions of the flux group on $C_0(\Ab_\Gamma)$. Furthermore, each triple forms a $C^*$-dynamical system $(C_0(\Ab_\Gamma),\alpha,\bar G_{\breve S,\Gamma})$. Finally, an action $\alpha$ of a particular flux group $\bar G_{\breve S,\Gamma}$ associated to a surface set $\breve S$ and a graph $\Gamma$, on the analytic holonomy $C^ *$-algebra $C_0(\Ab_\Gamma)$ defines a holonomy-flux cross-product $C^*$-algebra associated to the surface set and the graph.

\subsubsection{$O^*$- and $C^*$-algebras generated by unbounded and bounded operators}\label{subsec unbounded}

The well-known Gelfand-Na\u{\i}mark theorem states that any unital commutative $C^*$-algebra is isomorphic to the algebra of continuous functions on a compact topological space. But even more, the category of commutative unital $C^*$-algebras is dual to the category of topological spaces. The idea of the non-commutative Gelfand-Na\u{\i}mark theorem presented by Woronowicz and Kruszy\'{n}ski \cite{Woro82} is to give a correspondence between unital non-commutative $C^*$-algebras and non-commutative spaces. This duality is performed by the concept of a $C^*$-algebra, which is generated by a finite set of bounded and/or unbounded operators. 

Furthermore, the examples of Quantum Mechanics showed that a certain finite set of unbounded elements generates a $O^*$-algebra. The mathematical theory for $O^*$-algebras have been presented by Inoue \cite{Inoue} and Schm\"udgen \cite{Schmuedgen90}.
But these unbounded elements generate even a $C^*$-algebra in the sense of Woronowicz \cite{Woronowicz95,Woro91} or Schm\"udgen \cite{Schmuedgen90,Schmuedgenweb}. In the context of $C^*$-algebras the unbounded elements are not elements of the $C^ *$-algebra but they are affiliated. 

Thus the concept allows to define a huge number of different algebras generated by the quantum configuration and quantum momentum operators of a theory. Hence, for example in the article of Ashtekar and Isham \cite{AshIsh92} the holonomy $C^*$-algebra is generated by the Wilson functions $\tr(\ho(\gamma))$, which depends on a holonomy $\ho$ along a smooth loop $\gamma$  in the loop group\footnote{Notice that the loop group in this dissertation is not the loop group, which is often used in mathematics. The loop group in this context is a larger group.} $\LG(v)$ at a base point $v$, and some relations. This construction fits into the concept of Woronowicz. Some examples derived by Woronowicz can be compared with the algebra of Ashtekar and Isham. In particular, Woronowicz has analysed algebras given in the section before. The aim of Ashtekar and Isham was to find the spectrum of their commutative holonomy $C^*$-algebra. Until now the spectrum of the holonomy $C^*$-algebra of Ashtekar and Isham is not explicity known. The philosophy of Woronowicz is to define the continuous function algebra of a space vanishing at infinity by using a finite number of operators, relations and a appropriate norm. In \cite{KaminskiPHD} the idea of Woronowicz is used for the study of the holonomy $C^*$-algebra construction of Ashtekar and Isham. Their $C^*$-algebra is called the \textit{Wilson $C^*$-algebra} in this project \textit{AQV}. The reason for this is that this algebra is generated by the Wilson functions and relations among them. In \cite[Section 5]{KaminskiPHD} from the Wilson functions another $^*$-algebra will be derived by using a slightly different multiplication operation between Wilson functions. This algebra is completed to a new $C^*$-algebra, which is called the \textit{modified Wilson $C^*$-algebra}. Furthermore, a further $C^*$-algebra is obtained by using the algebra of almost periodic functions on the topological loop group $\LG(v)$. In this project \textit{AQV} this $C^*$-algebra is called \textit{smooth holonomy $C^*$-algebra}. All $C^*$-algebras are constructed independly from each other and, hence it is not apriori clear that, these algebras are isomorphic. A satisfactory description of these $C^*$-algebras is not given, such that the existence or non-existence of an isomorphisms cannot be proven until now. Summarising, the spectrum of the Wilson $C^*$-algebra is still unkown, but there are some other $C^*$-algebras constructable, which are similar to known algebras in mathematics.


\section{Quantum constraints, KMS-Theory and dynamics}\label{sec Iquant}

In this section a first short overview about the following issues are presented:
\begin{itemize}
 \item the classical and quantum system of constraints,
 \item the Dirac states,
 \item complete observables and
 \item KMS-states.
\end{itemize}

\subsection{The implementation of quantum constraints on algebras of Loop Quantum Gravity}

In the Hamilton formalism of General Relativity a set of constraints appear. They form a classical constraint algebra  on the hypersurface $\Sigma$. The classical variables are replaced by the quantisation map with Hilbert space operators or operator algebra elements. The structure of the constraint algebra is very difficult, since there is an infinite number of constraints. In \cite{Thiem96} and \cite{Thiemann97} Thiemann has analysed the quantum constraint algebra derived from holonomy and flux operators on a kinematical Hilbert space. In particular a formula for the quantum Hamilton constraint is presented. There is another idea to deal with a set of infinite classical constraints due to Thiemann. This is the Master constraint project, which has been developed in \cite{DittrichThiemann06,ThiemannPhoenix06}. The concept of the Master constraint can be reformulated in terms of quantum algebras and states. Originally in the LQG framework, the quantum constraints are usually given by Hilbert space operators. The implementation of the constraints is done for example by using rigging maps on Hilbert spaces. This can be found in the articles presented by Ashtekar, Lewandowski, Marolf, Mour\~{a}o and Thiemann \cite{AshLewMarolfMouraoThiem95}, or Giulini \cite{Giulini00} or Giulini and Marolf \cite{GiuliniMarolf99,GiuliniMarolfRed99}. In the operator algebra viewpoint used in this project \textit{AQV}, the constraints define particular states, which are contained in the state space of the algebra of quantum variables. It is required that this algebra contain the set of constraints (or at least the constraints are affiliated to the algebra). A state that implements in this sense the constraint is called a \textit{Dirac state}. In particular these states are invariant under automorphisms of the algebra, which are derived from the constraints. Furthermore, the time avarage of an operator is defined as a suitable state on the algebra. In LQG a contrary viewpoint is often used, the time avarage is given by an operator $T$ on a Hilbert space $\HS$. 

The implementation of constraints in a classical theory of gravity has been studied by many authors. In this project \textit{AQV} the work of Dittrich \cite{Dittrich05,Dittrich2006} will be focused. In her work the Dirac formalism has been used to perform a set of classical constraints that defines a contraint algebra, which is equipped with certain Poisson brackets on the phase space. Physical or Dirac observables are suitable phase space functions implemented by some particular Poisson brackets. In LQG approach to quantum gravity the classical Hamiltonian, which implements the dynamics of the system, is a constraint of the classical system, too. Therefore, the physical observables do not evolve with respect to this Hamiltonian. The evolution of a physical observable has to be related to a physical freedom of the system. This corresponds to a choice of a so called clock variable. In this project \textit{AQV} these concepts are introduced in the next sections. 

\subsubsection{The classical hypersurface deformation constraint algebra} 

Following Thiemann \cite{Thiemann01,Thiemann2002,ThiemannPhoenix06} in the classical theory the manifold 
\(M=\Sigma\times \R\) with hypersurface metric $q=(q_{ab})$ the following constraints occur 
\beqs \cdot\qquad&C_b(x)\text{ ... the spatial diffeomorphism constraint, }\\[3pt] 
\cdot\qquad&C(x)\text{ ... the Hamilton constraint and}\\[3pt]
\cdot\qquad&\mathcal{G}(x)\text{ ... the Gauss constraint}.\\[3pt]
\eqs for every $x\in\Sigma$.
The set of smeared constraints contains in particular all linear combinations of the smeared constraints:
\beqs\cdot\qquad&\VC(\VN)=\int_\Sigma d^3x N^a(x)C_a(x)\text{ ... the smeared spatial diffeomorphism constraint }\\ 
\cdot\qquad&C(N)=\int_\Sigma d^3x N(x)C(x)\text{ ... the smeared Hamilton constraint }\eqs which satisfy the following relations
\beq &\{\VC(\VN),\VC(\VN^\prime)\}=\kappa \VC(\LD_{\VN}\VN^\prime)\\ 
&\{\VC(\VN),C(N^\prime)\}=\kappa \VC(\LD_{\VN}N^\prime)\\ &\{C(N),C(N^\prime)\}=\kappa \int_\Sigma d^3x(N_{,a}N^\prime-NN^\prime_{,a})(x)q^{ab}(x)C_b(x)
\eq
for a constant $\kappa$, $\LD_{\overrightarrow{N}}\overrightarrow{N}^\prime$ is the Lie derivative for a vector valued function $\overrightarrow{N}^\prime$ on $\Sigma$ and $\LD_{\overrightarrow{N}}N$ is the Lie derivative for a scalar valued function $N$ on $\Sigma$. This set form the \textit{classical hypersurface deformation constraint algebra}.

Summarising the \textit{algebra of hypersurface quantum constraints} contains the quantum analogues of the classical spatial diffeomorphism constraints, the classical Hamilton constraints and the classical gauge constraints, and these quantum operators satisfy some certain relations.

\subsubsection{The quantum analogues of the classical Thiemann Master constraint, Dirac and complete observables}\label{subsubsec MasterHamiltonconstr}
\paragraph*{The issue of quantum constraints\\}\hspace{5pt}

In the following considerations the ideas of Thiemann \cite{ Thiemann01, Thiemann2002, ThiemannPhoenix06,Thiembook07} are reviewed and reinterpreted in the language of operator algebras. In Loop Quantum Gravity usually the configuration and momentum variables and the diffeomorphism, Gauss and Hamilton constraints are implemented as symmetric closed operators on a Hilbert space. In LQG \cite{AshLew_Back04,Thiemann01,Thiembook07} it has been possible to construct a Hilbert space such that the quantum configuration and momentum variables and the Gauss constraint are Hilbert space operator. The Hamilton constraint can only be imposed partly. In general the idea is to impose the quantum analogues of these constraints as one Master constraint, a set of constraint operators or an algebra of quantum constraints. This is studied in the next paragraphs.

The first approach is due to Thiemann \cite{ThiemannPhoenix06}, where he has proposed to consider only one self-adjoint positive operator $\MM$ instead of a set of constraints or an algebra.

The basic idea of Thiemann has been to replace a system of infinitely many constraints\\ $\mathfrak{C}(x)=(-q^{ab}C_aC_b)(x)+C(x)^2=0$ for every $x\in\Sigma$ by 
a single Master constraint $\MM$ formally given by
\beq \MM=\frac{1}{2}\int_\Sigma \dif^3x \frac{\mathfrak{C}(x)}{\sqrt{det(q)(x)}}\eq
where $q_{ab}$ is the spatial metric.

In the following paragraphs a more general Master constraint is studied. Let $J$ be a discrete finite index set.  Consider the Hilbert space $\HS^I_\MD=L^2(X^I_\MD,\mu^I_\MD)$, where $I\in J$ and $X^I_\MD$ is a Borel subset of a metrizable phase space $\MD$ and $\mu_\MD$ a Borel measure thereon. Let $\Sigma$ be a (metrizable) space and $X_\Sigma$ a Borel subset of $\Sigma$ with Borel measure $\nu_\Sigma$. Then for every $x\in\Sigma$ the constraint $C_I(x)$ depend also on the momentum space $\MD$, hence consider $\MD\ni m\mapsto C_I(x)[m]\in\CB$ such that $C_I(x)\in C^\infty(\MD)$. Then it is assumed that $C_I(x)$ is a multiplication operator acting on $\HS^I_\MD$. 

Therefore, the quantum Master constraint is a symmetric operator acting on the Hilbert space $\HS_\MD$. The operator is also denoted by $\MM$ and is defined by
\beqs \MM(m)&:= \int_{X_\Sigma}\dif\nu_\Sigma(x)  \sum_{I}C_I(x)[m]^*  C_I(x)[m]\qquad\text{ such that }\\
\langle \psi, \MM\phi\rangle_{\HS_\MD}& =\int_{X_\Sigma} \dif\nu_\Sigma(x) \sum_{I}\left\langle C_I(x)[m] \psi(m),  C_I(x)[m]\phi(m)\right\rangle_{\HS_\MD^I}
\eqs holds on a suitable dense domain. Assume that $\MM$ is positive, unbounded and essentially self-adjoint or positive, bounded and symmetric on $\HS_\MD$.
Then the constraint condition on $\MD$, which is given by
\beqs C_I(x)[m]=0\text{ }\forall x\in X_\Sigma^I,\eqs
is reformulated by the relation $\la\psi,\MM\psi\ra=0$ for all $\psi\in\HS_{\phys}$. This relation is called the \textit{Master constraint relation} in analogy to Thiemann. Therefore, there is a condition for the Hilbert space to be a \textit{physical Hilbert space}:
\beqs \HS_{\phys}:=\{\phi\in\HS_\MD: \MM\phi=0\}
\eqs for the Hilbert space operator $\MM$ with $0\in\sigma_{\disc}(\MM)$ (discrete spectrum). 

In the project \textit{AQV} the ideas are reformulated in the language of operator algebras.  Let $\Alg$ be an appropriate $C^*$-algebra of quantum variables and assume that $\Alg$ is a non-degenerate $C^*$-subalgebra of $\LD(\HS_\MD)$. 
Furthermore, let $\omega$ a state on $\Alg$. 
In the following paragraphs different constraints will be studied. 

Assume first that, the single exponentiated unitary Master constraint $\exp(i\MM)$ is contained in the multiplier algebra of the algebra $\Alg$. 
Then the Master constraint condition is replaced by the condition that the state $\omega$ corresponding to the GNS-representation of $\MM$ on the Hilbert space $\HS_\MD$ is a Dirac state. The Dirac state space is defined by
\beqs \Ss_D&:=\{\omega\in\Ss(\Alg): \pi_\omega(\exp(it\MM))\Omega=\Omega\text{ }\forall t\in\R\} \\
&=\{\omega\in\Ss(\Alg): \omega(\exp(it\MM)O)=\omega(O)=\omega(O\exp(it\MM))\quad\forall  O\in\Alg\text{ and }\forall t\in\R\} \\
\eqs whenever $(\pi_\omega,\HS_\omega,\Omega)$ is a GNS-representation associated to $\omega$. If the Master constraint $\MM$ is contained in $\Alg$, then $\omega$ is a Dirac state if $\omega(\MM)=0$ holds.

On the one hand in the previous paragraphs $\MM$ has been assumed to be a Hilbert space operator. This corresponds to a point-norm continuous one-parameter group $t\mapsto \alpha_t(\MM)$ of automorphisms on $\Alg$ that satisfies 
\[i\bra \MM,O\ket=\lim_{t\rightarrow 0}\frac{\alpha_t(\MM)(O)-O}{t}\]
Then a state is called \textit{$\R$-invariant} with respect to the automorphism group $t\mapsto \alpha_t(\MM)$ on $\Alg$, if
\beqs \omega\circ\alpha_t(\MM)=\omega\eqs holds for all $t\in\R$.
The set of $\R$-invariant states is denoted by $\Ss^{\alpha}$.

In analogy to Thiemann a bounded Hilbert space operator $O$, which satisfy
\beq\label{eq Masterrel1} \langle \psi, \left[[O,\MM],O\right]\psi\rangle =0\text{ for all } \psi\in\HS_{\phys}
\eq is called a \textit{weak Dirac observable}. If 
\beq\label{eq Masterrel2} \langle \psi, [O,\MM]\psi\rangle =0\text{ for all } \psi\in\HS_{\phys}
\eq holds, then $O$ is called a \textit{strong Dirac observable}.
In the following paragraph these objects are replaced. 

The commutativity condition \eqref{eq Masterrel2} is substituted by the condition for the state to be $\R$-invariant with respect to the automorphism group $t\mapsto \alpha_t(\MM)$ on $\Alg$. In particular Dirac states of the algebra $\Alg$ of quantum observables are $\R$-invariant states with respect to this automorphism group. Since it is assumed that $\MM$ is an element of $M(\Alg)$, the automorphisms $\alpha_t(\MM)$ are inner. In general a covariant representation of $(\Alg,\R,\alpha(\MM))$ in $\LD(\HS)$ is given by a pair $(\Phi_M,U_t(\MM))$ such that $U_t(\MM):=\exp(it\MM)$. Then the sets $\Ss^{\alpha}$ and $\Ss_D$ coincide. Consequently, all elements of the algebra $\Alg$ of quantum variables satisfy
\beq\label{eq cond commrel}\omega(i \bra O,\MM\ket)=\lim_{t\rightarrow 0}\frac{\omega(\exp(it\MM)O)-\omega( O\exp(it\MM))}{t}=0
\eq for every Dirac state $\omega$ of $\Alg$. A state independent formulation is the following. 
The  set $\Alg^\alpha$ of observables in $\Alg$ such that $\alpha_t(\MM)(A)=A$ for all $t\in\R$ is called the \textit{algebra of generalised strong Dirac observables} in this project. Then the condition $\omega(\bra O,\MM\ket)=0$ is fulfilled for every element $O$ in $\Alg^\alpha$ and every state $\omega$ in the state space of $\Alg$ (and where the convention $0/0=0$ is assumed).
 
For generality assume that $t\mapsto \alpha_t(\MM)$ is a one-parameter group of automorphisms on $\Alg$. Thiemann \cite{Thiemann01} has proposed an ergodic-mean operator on a Hilbert space. This will be generalised to arbitrary operators. Since in the project \textit{AQV} the states are focused, the following object will be used. Let $O$ be an element of the $C^*$-algebra $\Alg$ of quantum variables and let $\omega$ be a certain state on $\Alg$ with GNS-representation $(\HS_\Sigma,\pi_\Sigma,\Omega_\Sigma)$. Then a new state on $\Alg$ can be defined
\beq\label{eq timeavarage} \omega_\MM(O)&:=\limT \frac{1}{2T}\int_{-T}^T\dif t \text{ } \langle \Omega_\Sigma,\pi_\Sigma(\alpha_t(\MM)(O))\Omega_\Sigma\rangle\text{ for }(\pi_\Sigma,\Omega_\Sigma,\HS_\Sigma)\text{ GNS-triple assoc. to }\omega\\
&= \limT \frac{1}{2T}\int_{-T}^T\dif t  \text{ } \omega(\alpha_t(\MM)(O))
\eq such that $\omega_\MM\circ \alpha_t(\MM)=\omega_\MM$ for all $t\in\R$. Consequently, $\omega_\MM$ is a Dirac state and $O$ is a strong Dirac observable. Note that the right side of equation \eqref{eq timeavarage} is not necessarily well-defined and finite for all states on the algebra. In particular if $\omega$ is a density matrix state and a KMS-state for $\alpha_t(\MM)$, then the state $\omega_\MM$ defined by \eqref{eq timeavarage} exists, is invariant under the automorphism $\alpha$ and is a density matrix state, too. In general, if the state $\omega_\MM$ exists, then the state is called \textit{time average}. The remarkable properties of the time avarage, which is constructed from the KMS-state $\omega_\MM$, is one reason for the study of KMS-states in the project \textit{AQV}. But, often the state $\omega$ is not suitable. In these cases, the following operator can be defined 
\beqs
E(O):=&= \limT \frac{1}{2T}\int_{-T}^T\dif t  \text{ }\alpha_t(\MM)(O)
\eqs in the operator norm-limit and is called the \textit{time avarage operator}. Note that $\dif t  \text{ }\alpha_t(\MM)(O)$ is a positive operator-valued measure. This implies that a weight\footnote{A weight is a positive linear functional on the algebra, which is not necessarily normalizable.} $\tilde\omega_\MM$ on $\Alg$ is defined by
\beqs \tilde\omega_\MM(O)= \limT \frac{1}{2T}\int_{-T}^T\dif t \text{ } \omega(\alpha_t(\MM)(O))\in\bra 0,\infty\ket
\eqs for an arbitrary state $\omega$ on $\Alg$.

Summarising, for a Master constraint operator, which is contained in the algebra $\Alg$ of quantum observables in Loop Quantum Gravity, the time avarage construced from a KMS-state of $\Alg$ is an $\alpha$-invariant state and $\MM$ is imposed as a constraint on this state.

The situation is more difficult, if a set $\breve C:=\{C_J^*C_J\}$ of unbounded closed constraint operators $C_J$, which are not contained in a general $C^*$-algebra $\Alg$ of quantum observables, replaces the Master constraint. Assume that every element $Z_C$-transform defined by $C_J^*C_J$ is contained in the multiplier algebra of $\Alg$, and that $C_J^*C_J$ is essentially self-adjoint. Then the Dirac state space is given by
\beqs \Ss_D&:=\{\omega\in\Ss(\Alg):\pi_\omega(C^*_JC_J)\Omega_\omega=0\quad\forall C_J^*C_J\in\breve C\} \\
&=\{\omega\in\Ss(\Alg):\omega(C^*_JC_J)=0\quad\forall C_J^*C_J\in\breve C\}
\eqs
The \textit{set of strong Dirac observables} is given by the weak relative commutant 
\beqs \OD_D^s:= \{O\in\Alg: \omega(\bra O,C_J^*C_J\ket) =0\quad \forall C^*_JC_J\in\breve C\text{ and }\forall\omega\in\Ss_D\}\eqs 
and the \textit{set of weak Dirac observables} is given by 
\beqs \OD_D^w:= \{O\in\Alg: \omega(\bra O,\bra O, C_J^*C_J\ket\ket) =0\quad\forall C^*_JC_J\in\breve C\text{ and }\forall\omega\in\Ss_D\}\eqs 
The bracket $\bra O,C_J^*C_J\ket$ for a fixed $C_J^*C_J\in\breve C$ defines a $^*$-derivation $\delta_{C_J}(O):= \bra O,C_J^*C_J\ket$ on $\Alg$. Consequently, one can redefine the set $\OD_D^s$ by
\beqs \OD_D^s= \{O\in\Alg: \omega(\delta_{C_J}(O)) =0\quad\forall C_J\in\breve C\text{ and }\forall\omega\in\Ss_D\}\eqs

Now, a more general concept is introduced, which replaces the notion of weak and strong Dirac observables.
If products of constraints are used, then the Dirac states have to be analysed once more. Hence for generality assume that, $\breve C$ forms an algebra of quantum constraints and $\breve C$ is contained in the multiplier algebra $M(\Alg)$. 
Set \beqs N_\omega:=\{A\in \Alg: \omega(A^*A)=0\}\eqs 

Define the set $\ND$ to be the closed left and right ideal generated by $\breve C$. Then for example $AC$, $A^*C^*$, $CA$ and $C^*A^*$ are elements of $\ND$. 
Denote the closure of all linear combinations of elements of $\ND$ and $\breve C$ by $\DD$. Then for a constraint $C$ in $\breve \CD$ it is in particular true that $C\in\DD$, $C^*A^*\in\DD$, $\bra A,C\ket\in\DD$ and $\bra\bra A,C\ket,A\ket\in\DD$. 

Redefine the \textit{Dirac state space} by 
\beqs\Ss_D:=\{\omega\in\Ss(\Alg):\pi_\omega(D)\Omega_\omega=0\quad\forall D\in\DD\}
\eqs Then $\DD\subset N_\omega$ whenever $\omega\in\Ss_D$. 

Then in this project the \textit{set of Dirac observables} is given by 
\beqs \OD_D:=\Alg / \DD
\eqs
The set $\OD_D$ forms in particular a $^*$-algebra, which can be hopefully completed to a $C^*$-algebra.

Consider for every $J$ the one-parameter group $\R\ni t\mapsto \alpha_t(C_J^*C_J)\in\Aut(\Alg)$ of automorphism such that $C_J^*C_J$ defines the (infinitesimal) generator of this group.
Then the set $\bigcap_J\Ss^{\alpha_J}$ of all states of $\Alg$, which are invariant under all automorphism groups $\R\ni t\mapsto\alpha_t( C^*_JC_J)\in\Aut(\Alg)$ for all constraints in $\breve C$, is not contained in the set $\Ss_D$ of Dirac states. It is only true that a Dirac state is invariant under every automorphism  $\alpha_t( C^*_JC_J)$. Consequently, the state $\omega_\MM$, where $\MM$ is replaced by a constraint in $\breve C$, defined by \eqref{eq timeavarage} need not be a Dirac state. Hence, a Dirac state is a more general concept than states, which are invariant under automorphisms given by the constraints. 

There is a problem, if the constraints or the exponentiated constraints are not contained in the algebra $\Alg$ or the multiplier algebra of $\Alg$. Then in some cases the quantum constraint is affiliated to a larger algebra of quantum configuration and momentum variables. Then similar investigations can be made with respect to this larger algebra. Note in this situation the algebra of quantum constraints can be replaced by the algebra generated by all $Z$-transformations of the constraints $C_J$. The $Z$-transform of an operator $C_J$ has been given in \cite{Schmuedgenweb} by $C_J(\idf + C_J^*C_J)^{-1/2}$. These objects are in particular elements of the multiplier algebra of $\Alg$ if the constraints $C_J$ are affiliated operators with $\Alg$.

On the other hand, a more general implementation of quantum constraints in terms of multipliers is the following. If the non-unital $C^*$-algebra $\breve \CD$ of constraints is a concrete $C^*$-algebra of bounded operators on a separable Hilbert space $\HS$, then the multiplier algebra $M(\breve \CD)$ of the $C^*$-algebra $\breve \CD$ defines the $C^*$-algebra of quantum observables. In Loop Quantum Cosmology, for example, this characterisation can be used. 

In general for the Loop Quantum Gravity approach the algebra, which is generated by the quantum constraint operators, has not been understood completely. The project of \textit{Algebras of Quantum Variables (AQV) for LQG} will change this. The algebras of quantum variables have to be chosen such that the operators derived from the quantum constraints are
\begin{itemize}
 \item elements of (or affilliated with) the algebra of quantum variables, or
 \item elements of the multiplier algebra of this algebra.
\end{itemize}

\paragraph*{Partial and complete quantum observables\\}\hspace{5pt}

Apart from the issue of the implementation of quantum constraints, the issue of partial and complete quantum observables has to be analysed. Assume that the Master constraint operator is contained in the multiplier algebra of $\Alg$ and there are no other constraints. Let $\MM$ be the Master constraint, which implements the dynamics of the gravitational system and, therefore, defines an one-parameter group of automorphisms $\tau\mapsto\alpha_{\MM}(\tau)$, which is defined by
\beq\label{eq limitexistence} \alpha_{\MM}(\tau)(A)
&:=\limN \sum_{n=1}^N\frac{\tau^n}{n!}\delta_{\MM}^n(A)\text{ for }\delta_{\MM}^n(A):=\bra A,\MM\ket_n=\bra ...\bra \bra A,\MM\ket,\MM\ket,...,\MM\ket
\eq for all $A\in\Alg$ such that the limit exists in norm-topology.

Finally, let the state $\omega_\MM$ of the quantum algebra $\Alg$ be a Dirac state. Then in particular, this state is invariant under the one-parameter group of automorphisms, which is given by $s\mapsto\alpha_{\MM}(s)$. Therefore, the physical or Dirac observables do not evolve with respect to this Hamiltonian. 

The \textit{set of complete quantum observables} is defined by
$A_{\tau}=\alpha_{\MM}(\tau)(A)$ running over all $A\in\OD_{D}$ such that the limit \eqref{eq limitexistence} exists in norm-topology. In particular, the set \beqs \OD^{\alpha_{\MM}}_{D}:=\{A\in\OD_{D}:\alpha_\MM(t)(A)=A\quad\forall t\in\R\}\eqs forms an algebra. This algebra can be completed to a $C^*$-algebra, which is called the \textit{$C^*$-algebra of complete quantum observables} in this project.

The evolution of a physical observable has to be related to a clock variable $T$ of a physical freedom of the system. Let $T$ be an element of the $C^*$-algebra  $\OD_{D}$ or the multiplier algebra of $\OD_{D}$.  
Furthermore, let $H_T$ be the Hamiltonian of the clock observable $T$ and assume that $t\mapsto\alpha_{H_T}(t)$ is a one-parameter group of automorphisms on $\OD_{D}$. 
Then the \textit{time-of-occurence-of-an-event operator} is defined similarly to the operator, which has been introduced by Fredenhagen and Brunetti in \cite{BrunettiFredenhagen02}, and is given by the operator
\beqs E(A):= \int_{\tau}\dif t \text{ }\tilde T^{-\nicefrac{1}{2}}\alpha_{H_T}(t)(A^*A)\tilde T^{-\nicefrac{1}{2}}
\eqs Note that $\tilde T$ is derived from the clock operator $T$ and have to be chosen suitably. Furthermore, the expectation value of an observable $A^*A$ contained in $\OD^{\alpha_{\MM}}_{D}$, if the clock $T$ measures a time endurance $\tau$, is given by 
\beq\label{eq exptimeocc} W_T(A^*A):=\int_{\tau}\dif t\text{ }\omega_\MM(\tilde T^{-\nicefrac{1}{2}}\alpha_{H_T}(t)(A^*A)\tilde T^{-\nicefrac{1}{2}})\eq Then $W_T$ is a linear functional, but it is not necessarily a state on the $C^*$-algebra of complete quantum observables. In general, it is a weight on this $C^*$-algebra. If it is a state, then it is called the \textit{expectation of the time of occurence of an event}. 

In the context of the thermal time hypothesis presented by Rovelli and Connes in \cite{ConnesRovelli} the one-parameter group of automorphism $t\mapsto \alpha_{H_T}(t)$ is the modular group and $\omega_\MM$ is the thermal equilibrium with respect to the thermal time $t$, which is given by the Hamiltonian of the clock.   
The elements of $\Alg$ are called \textit{partial quantum observables} in the project \textit{AQV}. 

Summarising, the complete or partial quantum observables are certain  elements of the $C^*$-algebra of quantum variables. In particular the complete quantum observables are assumed to define a certain $C^*$-subalgebra of the $C^*$-algebra of quantum variables. The constraints are imposed on a Dirac state and the complete quantum observables are assumed to be Dirac observables. Moreover a Dirac state, a clock operator and a clock Hamiltonian define a state or a weight on the $C^*$-algebra of complete quantum observables.

There exists several one-parameter groups of automorphisms on the $C^*$-algebra of complete observables. The modular automorphism group, which is related to a physical time evolution, is defined by the quantum clock Hamiltonian. The ideas are presented in more detail in subsection \ref{subsec KMStheory}. Clearly the concept has to be further generalised if a set of clocks is used.

In LQG the algebra of quantum variables that contains the quantum constraint and quantum clock operators is very complicated and has yet not been developed completely. Moreover, the algebra of complete quantum observables has not been derived from the full algebra of quantum variables so far. 

\subsection{KMS-Theory in Generally Covariant Theories }

In the Hamiltonian formulation of Gravity the dynamical Hamiltonian is a constraint. A preferred time flow such that the physical observables evolve with respect to this single time parameter is related to the concept of clocks. The idea of Modular Theory is to encode the time flow in a one-parameter group of automorphisms, which depends on the thermal state of the system. Hence, one can speak about thermal time. This concept is analysed in the article \cite{ConnesRovelli} of Connes and Rovelli in the context of general covariant quantum theories and Modular Theory for von Neumann algebras.

Mathematically, for a von Neumann algebra $\MK$ there exists the modular automorphism group $t\mapsto \alpha^\omega_t\in\Aut(\MK)$ associated to a faithful and normal state $\omega$. This automorphism group is unique up to inner automorphism and is independent of the choice of the state. Moreover, the modular automorphism group is used for the study of type III factors of von Neumann algebras. 

More precisely the KMS-theory contains the following objects. In the GNS representation $(\HS,\pi,\Omega)$ associated to the state $\omega$ there exists a unitary one-parameter group $t\mapsto \Delta^{it}_\omega\in\LD(\HS)$ such that $\pi(\alpha^\omega_t(M))=\Delta^{it}_\omega\pi(M)\Delta^{it}_\omega$ for all $M\in\MK$. The operator $\Delta_\omega$ is called the modular operator. There exists the modular generator $K_\omega:=\log \Delta_\omega$, which is the generator of the automorphism group $t\mapsto\alpha^\omega_t$. Furthermore, there exists a anti-linear isometry $J$ in $\HS$ and an isomorphism $\gamma:\pi(\MK)\rightarrow\pi(\MK)^\prime$ such that $\gamma(\pi(M))=J\pi(M)J$ for all $M\in\MK$. The operator $J$ is called the modular conjugation. The modular automorphism is the only one-parameter automorphism group satisfying the KMS-condition w.r.t. the state $\omega$ at inverse temperature $\beta=1$. The KMS-condition states that $\omega\circ\alpha_t^\omega=\omega$ and for all $M,N\in \MK$ there exists a map $F_{M,N}:\R\times\bra 0,\beta\ket\rightarrow \ket 0,\beta\bra$ such that $F_{M,N}$ is holomorphic on $\R\times \bra 0,\beta\ket$, $F_{M,N}$ is bounded continuous on $\R\times\bra 0,\beta\ket$, $F_{M,N}(t)=\omega(\alpha^\omega_t(B)A)$ and $F_{M,N}(i\beta+t)=\omega(A\alpha_t^\omega(B))$ for all $t\in \R$.

Consequently, the dynamics defined by the modular operator $\Delta_\omega$, and a KMS-state $\omega$ are intrinsic objects of the von Neumann algebra. Physically, the equilibrium thermal state $\omega$ and the modular automorphism group contains all information about the dynamics of the system. In particular, the information about the Hamiltonian, which is the generator of the automorphism group. An overview abot these structures can be found in the book \cite{BratteliRobinsonB1} of Bratteli and Robinson and for a detailed lecture refer to the books of Takesaki \cite{TakesakiI_2003} and \cite{TakesakiII_2003}. 

In the fundamental article \cite{Emch1981} of Emch the role of KMS-Structures and a quantisation of a classical Poisson system is analysed. He considered the von Neumann algebra $\MD$ generated by the unitary Weyl elements $W(x,p)$, which satisfy the canonical commutator relation and which are based on the phase space $\R^{2n}$. Then he showed that for every faithful normal state $\omega$ on $\MD$ there is a cyclic and separating vector $\Phi$ in a Hilbert space $\HS$ such that $\omega(W)=\la \Phi, W\Phi\ra_{\HS}$ for all $W\in\MD$. Furthermore, every faithful normal state $\omega$ on $\MD$ satisfies the KMS-condition and, hence, the modular objects are constructable.

In Quantum Field Theories the Tomita-Takesaki theory lead to a surprising duality between geometric objects on Minkowski spacetime and the modular automorphism group on the algebra of local observables. For example Brunetti, Guido and Longo \cite{BrunettiGuidoLongo} have shown that certain one-parameter subgroups of the Poincar\'{e} group are related to certain modular groups constructed via Tomita-Takesaki theory. In particluar, the Bisognano-Wichmann theorem relates the Lorentz boosts on restricted wegde regions in Minkowski spacetime to unitaries, which are defined by the modular generator. Consequently, the boosts implement the dynamical evolution of free fields in Minkowski spacetime. The modular involution $J$ implements the spacetime reflection about the edge of the wedge, along with a charge conjugation. Note that, the algebra of observables is restricted to a certain subalgebra associated to wedges and the full Poincar\'{e} invariance of the representation is broken. Then Brunetti, Guido and Longo assumed that the subgroup of Lorentz boosts is implemented as a covariant representation on the $C^*$-dynamical system consisting of the algebra of local observables restricted to wedges, the boosts and the automorphisms, which implement the boosts. This representation is also called the thermal representation $(\pi_\beta,\HS_\beta,\Omega_\beta)$. The self-adjoint thermal Hamiltonian $H_\beta$ is the generator of the unitary group $U_\beta(t):=\exp(-i\beta H_\beta)$ such that $\la\Omega_\beta,A\exp(i\beta H_\beta)B\Omega_\beta \ra =\la \Omega_\beta,BA\Omega_\beta\ra$ for elements $A,B$ contained in a suitable dense subset of the quantum algebra.
Indeed, the vacuum representation is not related to a KMS-state at a finite temperature. A consequence of the Connes cocycle theorem is that there is only one thermal state and automorphism group (up to inner automorphisms) of the quantum von Neumann algebra. This implies that there is also only one preferred time evolution of the physical system. In general modular groups on von Neumann algebras in the QFT framework are considered by Borchers in \cite{Borchers95}. A short summary over Tomita-Takesaki Theory in QFT is presented by Summers in \cite{Summers07}.

There is also a modular theory on $C^*$-algebras such that KMS-states and modular objects can be defined. This lead physically to the concept that the $C^*$-algebra of quantum operators and the modular automorphism are important objects for the definition  of a theory of quantum gravity. Clearly, for this viewpoint it is assumed that the algebra of constraints is a subset of (or are affiliated with) the algebra of quantum variables.

In the framework of Loop Quantum Gravity KMS-sates and modular objects have not been studied until now. The questions that arise are the following:
\begin{itemize}
 \item Which automorphism group of the algebra of quantum variables is a candidate for the modular automorphism group?
 \item Which Hamiltonian is required to be the generator of the modular automorphism group?
\end{itemize}
The answers will depend on the choice of the algebra of LQG. In this project it is shown that, already for some simple automorphisms of the known $C^*$-algebras of quantum variables, there exists no KMS-states. Furthermore, the von Neumann algebra generated by holonomies and fluxes is not suitable. Consequently, the study of new algebras is necessary, if one would like to explore KMS-theory in LQG. 

\section{The quantum variables of Loop Quantum Gravity}\label{subsec algLQG}

In the last section the reasons for a study of different algebras of quantum variables have been collected. The ideas used in the examples of Quantum Mechanics, which have been presented in subsection \ref{subsec QM},  are used for a construction of algebras in LQG. Furthermore, the choice of the quantum analogues of the classical configuration and momentum variables of the theory of LQG allows to configure various algebras, too.

There are many different degrees of freedom for a construction of an algebra of quantum variables in the context of LQG. For example, different algebras arise if the following choices are made:
\begin{enumerate}
\item\label{listofoptions1} The configuration space changes if loops or paths are chosen to be smooth or semi-analytic.
\item\label{listofoptions2}  The algebra of quantum configuration variables contains different functions, which depend on holonomies along loops or paths.
\item\label{listofoptions3}  Different multiplication operations and involutions define different $^*$-, $O^*$- or $C^*$-algebras of quantum variables.
\item\label{listofoptions4}  The algebras of quantum variables can be completed with respect to different norms such that different Banach $^*$-algebras or $C^*$-algebras arise.
\item\label{listofoptions5}  The group- or Lie algebra-valued quantum flux operators are (or are not) contained in an algebra of quantum variables. If they are not contained in a $C^*$-algebra of quantum variables, then they are assumed to be affiliated with this algebra.
\item\label{listofoptions7}If inductive families of graphs are considered, then the $C^*$-algebras can be defined as inductive limit $C^*$-algebras of inductive families of $C^*$-algebras of quantum variables restricted to graphs.  There is also an inductive limit derived from an inductive family of finite graph systems. Clearly, there exists many other inductive limit $C^*$-algebras, which can be obtained for inductive families of other $C^*$-algebras restricted to other of particular sets paths.
\item\label{listofoptions6} Other quantum operators, for example the quantum curvature or a generalised holonomy map along paths, are not contained in and are not affiliated with the usual algebras of quantum variables in LQG. Then new algebras generated among other operators by these new quantum constraints exist.
\end{enumerate}

A short summary over the different algebras obtained by different choices is given in this section. The complete development can be found in the PhD thesis \cite{KaminskiPHD} or in several subsequent articles \cite{Kaminski1,Kaminski2,Kaminski3,Kaminski4,Kaminski5}. For an introduction into the basic quantum variables in LQG refer to \cite[Chapter 2]{Kaminski1} or \cite[Chapter 3]{KaminskiPHD}. A short overview is given in the next subsection.

\subsection{The quantum configuration variables: holonomies along paths}

The fundamental geometric objects for a theory of Loop Quantum Gravity are (semi-) analytic paths and loops that form graphs. In the project \textit{AQV} the following objects are often used. 

A \textit{graph} contains a finite set of independent edges. A set of edges is called \textit{independent} if the edges only intersect each other in the source or target vertices. A \textit{finite groupoid} is a finite set of paths equipped with a groupoid structure. The \textit{finite graph system} associated to a graph $\Gamma$ is given by all subgraphs of $\Gamma$. A \textit{finite path groupoid} associated to the graph $\Gamma$ is generated by all compositions of elements or their inverse elements of the set of edges that defines the graph $\Gamma$. Note that, an element of a finite path groupoid is not necessarily an independent path. Clearly, for all these objects there exists an ordering such that 
\begin{enumerate}
 \item\label{indfam1} an \textit{inductive family of graphs}
 \item\label{indfam2} an \textit{inductive family of finite path groupoids} and
 \item\label{indfam3} an \textit{inductive family of finite graph systems} can be studied.
\end{enumerate} 

Furthermore, a \textit{holonomy map} is a groupoid morphism from the path groupoid to the structure group $G$. If a graph is considered, then the holonomy map maps each edge of the graph to an element of the structure group $G$. For generality it is assumed that $G$ is a compact group. In \cite[Section 2.2.2]{Kaminski1} or \cite[Section 3.3.4]{KaminskiPHD} two ways of an identification of the holonomy map evaluated for a subgraph of $\Gamma$ with elements in $G^{\vert\Gamma\vert}$ will be presented. One distinguishes between the \textit{natural} or \textit{the non-standard identification of the configuration space} $\Ab_\Gamma$ with $G^{\vert\Gamma\vert}$. Recall that a subgraph of $\Gamma$ is a set of independent paths, which are generated by the edges of the graph $\Gamma$. In the natural identification these paths are decomposed into the (or the inverse) edges, which define the graph $\Gamma$. In the non-standard identification only graphs that contain only non-composable paths are considered. In both cases the holonomy maps evaluated on a subgraph $\Gp$ of $\Gamma$ are elements of $G^{M}$, where $M$ is the number of paths in $\Gp$. One obtains a product group $G^M$ for $M\leq \vert\Gamma\vert$, and which is embedded into $G^{\vert\Gamma\vert}$ by $G^M\times\{e_G\}\times ...\times \{e_G\}$. Hence, in both cases the holonomy evaluated on a subgraph of a graph $\Gamma$ is an element of $G^{\vert\Gamma\vert}$. In LQG \cite{AshLew93,AshLew94,Thiembook07} a holonomy map evaluated at the graph $\Gamma$ is an element of $G^{\vert\Gamma\vert}$, too.

The \textit{analytic holonomy $C^*$-algebra restricted to a finite graph system associated to a graph} is given by the commutative $C^*$-algebra $C(\Ab_\Gamma)$ of continuous functions on the configuration space $\Ab_\Gamma$ vanishing at infinity and supremum norm.

In the project \textit{AQV} the inductive limit $C^*$-algebra is constructed from an inductive family of $C^*$-algebras, which depend on finite graph systems. The reason is the following: Consider \textit{graph-diffeomorphisms} of the finite graph system associated to a graph $\Gamma$. These objects are pairs of maps and are presented in more detail in subsection \ref{subsec quantumconstraints}. For short such a pair consists of a bijective map from vertices to vertices, which are situated in the manifold $\Sigma$, and a map that maps subgraphs to subgraphs of $\Gamma$. Then there are actions of these graph-diffeomorphisms on the analytic holonomy $C^*$-algebra restricted to a finite graph system associated to the graph $\Gamma$. There is no well-defined action of these graph-diffeomorphisms on the analytic holonomy $C^*$-algebra restricted to a fixed graph in general. This can be verified as follows. Assume that $\Gamma:=\{\gamma_1,\gamma_2,\gamma_3\}$ is a graph and $\Gp:=\{\gamma_1\}$, $\Gpp:=\{\gamma_1\circ\gamma_3\}$ are subgraphs of $\Gamma$ . Then consider a graph-diffeomorphism $(\varphi,\Phi)$ such that $\Phi(\Gp)=\Gpp$. Now the action $\zeta_{(\varphi,\Phi)}$ on the analytic holonomy $C^*$-algebra restricted to the graph $\Gamma$, which is defined by 
\beqs (\zeta_{(\varphi,\Phi)}f_\Gamma)(\ho_\Gamma(\Gamma))=f_{\Phi(\Gamma)}(\ho_{\Phi(\Gamma)}(\Phi(\Gamma)))= f_{\Gppp}(\ho_{\Gppp}(\Gppp))
\eqs whenever $\Phi(\Gamma)=\Gppp=\{\gamma_1\circ\gamma_3,\gamma_2,\gamma_3\}$ is not well-defined. The reason is: $f_{\Gppp}$ is not an element of the analytic holonomy $C^*$-algebra restricted to the graph $\Gamma$ and $\Gppp$ is not a graph. If $\Phi(\Gamma)$ is a subgraph of $\Gamma$, then in particluar $f_{\Phi(\Gamma)}$ is an element of the analytic holonomy $C^*$-algebra restricted to the subgraph $\Phi(\Gamma)$. The analytic holonomy $C^*$-algebra restricted to the graph $\Gamma$ is a $C^*$-subalgebra of the analytic holonomy $C^*$-algebra restricted to the finite graph system associated to the graph $\Gamma$. An action of graph-diffeomorphisms is an automorphism of the analytic holonomy $C^*$-algebra restricted to finite graph system associated  to $\Gamma$. Summarising, the concepts of the limit of $C^*$-algebras restricted to finite graph systems, and actions of graph-diffeomorphisms on the holonomy $C^*$-algebra restricted to finite graph systems engage with each other.

Finally note that the inductive limit $C^*$-algebra of the inductive family of $C^*$-algebras $\{C(\Ab_\Gamma),\beta_{\Gamma,\Gp}\}$ defines the \textit{projective limit configuration space} $\Ab$. The inductive limit $C^*$-algebra $C(\Ab)$ is called the \textit{analytic holonomy $C^*$-algebra} in the project \textit{AQV}.

The idea of using families of graph systems is influenced by the work of Giesel and Thiemann \cite{ThiemGiesel} in the LQG framework. They have used particular cubic graphs instead of sets of paths in a groupoid and their inductive limit has been constructed from families of cubic graph systems. 
In the project \textit{AQV} the \textit{inductive limit Hilbert space} $\HS_\infty$ will be derived from the natural or non-standard identified configuration spaces, the Haar measure on the structure group $G$ and an inductive limit of finite graph systems. It will be assumed that, the inductive limit graph system only contains a countable set of subgraphs of an inductive limit graph $\Gamma_\infty$. This is contrary to the Hilbert space used in LQG literature \cite{Thiembook07}, which is given by the Ashtekar-Lewandowski Hilbert space $\HS_{\text{AL}}$. The Hilbert space $\HS_{\text{AL}}$ is manifestly non-separable, since the limit is taken over all sets of paths in $\Sigma$ and, hence over an infinite and uncountable set of all graphs. Clearly, the Hilbert space $\HS_\infty$ is constructed by using certain identification of the configuration space and the countable set of subgraphs. In this simplified formulation some important aspects of the theory can be studied. It is possible to generalise partly the results for the Ashtekar-Lewandowski Hilbert space. 

The \textit{classical configuration space} in the context of LQG and Ashtekar variables is the space of smooth connections $\breve\A_s$ on an arbitrary principal fibre bundle $P(\Sigma,G)$. In this project the quantum operator $\QD(A)$ of the infinitesimal connection $A$ is given by the holonomy $\ho$ along a path $\gamma$. The operator $\QD(A)$ is represented as a multiplication operator on the inductive limit Hilbert space $\HS_\infty$.

For a construction of a completely new algebra of quantum variables, which is derived from holonomies, fluxes and curvature, the setup of the configuration variables has to be changed. This will be described in section \ref{subsec Holgroupoid}.

\subsection{The quantum momentum variables: group-valued or Lie algebra-valued flux operators}\label{subsec fluxop}

In the project \textit{AQV} the quantum operator $\QD( E^i)$ of the classical flux $E^i$ is either a group- or Lie algebra-valued operator, which depend on a surface $S$ and a path $\gamma$ or a graph $\Gamma$.  The idea of this definition is the following: Consider a surface $S$ and a path $\gamma$ that intersets each other in the source vertex of $\gamma$ and the path lies below the orientated surface $S$. Let $\go$ be the Lie algebra of a compact connected (linear) Lie group $G$. The \textit{Lie algebra-valued quantum flux operator} $E_S(\gamma)$ is given by the value of a map $E_S: P\Sigma\rightarrow \go$ evaluated for a path $\gamma$ in the set $P\Sigma$ of paths in $\Sigma$. This definition does not coincide with the usual definition presented in LQG literature completely. In this project the flux-like variables introduced by Lewandowski, Oko\l{}\'{o}w, Sahlmann and Thiemann \cite{LOST06} are replaced and generalised to Lie algebra-valued quantum flux operators. The \textit{group-valued quantum flux operator $\rho_S(\gamma)$} are defined similarly by suitable maps $\rho_S:P\Sigma\rightarrow G$.

In general the idea is to obtain algebras, which are generated by 
\begin{enumerate}
 \item the group-valued quantum flux operators and the holonomies along paths in a graph, or
\item the group-valued quantum flux operators and functions depending on holonomies along paths in a graph, or
\item\label{item LA1} the Lie algebra-valued quantum flux operators and the holonomies along paths in a graph, or
\item\label{item LA2} the Lie algebra-valued quantum flux operators and functions depending on holonomies along paths in a graph.
\end{enumerate}
In the following algebras, which are generated among other operators by the Lie algebra-valued quantum flux operators, are presented. Therefore consider either \ref{item LA1} or \ref{item LA2} and the some certain canonical commutator relations.

The $\go$-valued quantum flux operator $E_S(\gamma)$ and the holonomy $\ho$ along a path $\gamma$ satisfy the canonical commutator relation, which is given by
\beq\label{eq ccr}  \bra E_S(\gamma),\ho(\gamma)\ket = \frac{\dif}{\dif t}\Big\vert_{t=0}\exp(tE_S(\gamma))\ho(\gamma)  -\ho(\gamma)E_S(\gamma)
\eq whenever $t\in\R$. Set 
\beqs E_S(\gamma)\ho(\gamma):= \frac{\dif}{\dif t}\Big\vert_{t=0}\exp(tE_S(\gamma))\ho(\gamma)  
\eqs 

Furthermore the \textit{right-invariant flux vector field} $e^{\overrightarrow{L}}$ is defined by
\beq\label{eq fluxcross}\bra E_S(\gamma),f_\Gamma\ket=e^{\overrightarrow{L}}(f_\Gamma)\eq where
\beq\label{CommRel1} e^{\overrightarrow{L}}(f_\Gamma)(\ho_\Gamma(\gamma)):=\frac{\dif}{\dif t}\Big\vert_{t=0} f_\Gamma(\exp(t X_S)\ho_\Gamma(\gamma))&\text{ for }X_S\in\go, \ho_\Gamma(\gamma)\in G, t\in\R
\eq whenever $f_\Gamma\in C^\infty_0(\Ab_\Gamma)$.

The quantum flux operator $E_S(\Gamma)$ is represented as the differential operator $\frac{\dif}{\dif t}\exp(tE_S(\gamma))$ on the Hilbert space $\HS_\Gamma$. The holonomies along paths or the functions depending on holonomies along paths are represented as multiplication operators on the Hilbert space $\HS_\Gamma$.

Until now, a suitable set of surfaces in $\Sigma$ and a path $\gamma$ in the finite path groupoid $\PD_\Gamma\Sigma$ are fixed. For a general situation the following maps are studied in \cite[Section 2.4 and 2.5]{Kaminski1},\cite{Kaminski3}, \cite[Section 3.4]{KaminskiPHD}:
\begin{enumerate}
 \item\label{maps1} a certain map $E_S:\PD_\Gamma\Sigma\rightarrow\go$ 
(refer to \cite[Definition 3.4.1]{KaminskiPHD}),
\item\label{maps2} a certain map $E_S:\PD_\Gamma\Sigma\rightarrow\E$ 
(refer to \cite[Definition 3.4.10]{KaminskiPHD}), where $\E$ is the enveloping algebra of $\go$,
\item\label{maps3} a certain map $\rho_S:\PD_\Gamma\Sigma\rightarrow G$ 
(refer to \cite[Definition 3.4.14]{KaminskiPHD}),
\item\label{maps4} a cetrain map $\rho_S:\PD_\Gamma\Sigma\rightarrow \ZD$ 
, where $\ZD$ denotes the center of the group $G$, and
\item\label{maps5} a certain map $\varrho:\PD_\Gamma\Sigma\rightarrow G$ 
(refer \cite[Definition 3.4.21]{KaminskiPHD}) and this map $\varrho$ is called \textit{admissible} in analogy to Fleischhack \cite{Fleischhack06}.
\end{enumerate}
Then the maps of the form $E_S$ given by \ref{maps1} (or \ref{maps2}) define a Lie algebra (or an enveloping algebra), which depends on a fixed path $\gamma$ in $\PD_\Gamma\Sigma$ and on surfaces in a suitable fixed surface set $\breve S$. Note that the surface set always contains at least one surface in $\Sigma$. This Lie algebra is called the \textit{Lie flux algebra associated to a surface set and a path}. The maps $\rho_S$ given by \ref{maps3} (or \ref{maps4}) define a group, which depends on the fixed path $\gamma$ and a suitable fixed surface set $\breve S$. This group is called \textit{flux group $\bar G_{\breve S,\gamma}$ associated to a surface set $\breve S$ and a path $\gamma$}. Clearly, for each suitable surface set there exist a flux group associated to this surface set. The maps  of the form $\varrho$ given by \ref{maps5} are used to define a more complicated structure. Furthermore, this concept generalises to holonomies of a graph $\Gamma$, which are maps from graphs to products of the structure group $G$. Then for example the \textit{flux group $\bar G_{\breve S,\Gamma}$ associated to a surface set and a graph} is defined in \cite[Definition 3.4.14]{KaminskiPHD}. 

Now, for the group-valued or the Lie algebra-valued quantum flux operators different actions on the configuration space will be explicitly considered in \cite[Section 3.1]{Kaminski1} or \cite[Section 6.1]{KaminskiPHD}. In particular the left, right and inner actions are studied independently from each other and are denoted by $L$,$R$ or $I$. Furthermore, only the maps \ref{maps4} and \ref{maps5} define groupoid morphisms by composition of the action $L$ (or $R$, or $I$) and the holonomy map. For an overview about which maps define groupoid morphisms consider \citetableB. Note that, using admissible maps (maps of the form \ref{maps5}) particular morphisms are defined. These morphisms are called \textit{equivalent groupoid morphisms} in analogy to Mackenzie \cite{Mack05} and are related to gauge transformations on the configuration space. The flux groups constructed from the maps \ref{maps3} and \ref{maps4}, the analytic holonomy $C^*$-algebra $C(\Ab_\Gamma)$ and the actions $L$, $R$ or $I$ define $C^*$-dynamical systems. If admissible maps are taken into account, the $C^*$-dynamical systems are very complicated. 

The starting point of Fleischhack's construction \cite{Fleischhack06} of an algebra has been the analysis of homeomorphisms on the projective limit configuration space $\Ab$. He has assumed that $G$ is a compact connected Lie group. The analytic holonomy algebra has been given by the unital commutative $C^*$-algebra $C(\Ab)$ and is represented on the Hilbert space $\HS_{\text{AL}}$ as multiplication operators. The Ashtekar-Lewandowski Hilbert space $\HS_{\text{AL}}$ is given by $L^2(\Ab,\mu_{\text{AL}})$, where $\mu_{\text{AL}}$ is a measure on $\Ab$. Measure preserving transformations are implemented by certain homeomorphisms on the configuration space $\Ab$ and they correspond to unitary operators on the Hilbert space $\HS_{\text{AL}}$. The Weyl algebra of Quantum Geometry, which has been introduced by Fleischhack, is generated by functions in $C(\Ab)$ and these unitaries. Hence, elements of the Weyl algebra are for example of the form $f$, $fU$ or $U$ if $f$ is an element of $C(\Ab)$ and $U$ is a unitary operator on the Hilbert space $\HS_{\text{AL}}$. On the other hand, homeomorphisms on the projective limit Hilbert space define automorphisms on the $C^*$-algebra $C(\Ab)$. In the project \textit{AQV} these automorphisms play a fundamental role. 

But for example the parameter group of automorphism, which is defined from arbitrary group-valued quantum flux operators $\rho_S(\gamma)$ for every surface $S$ and a fixed path $\gamma$ to the group of automorphisms, i.e. $\rho_S(\gamma)\mapsto\alpha(\rho_S(\gamma))\in\Aut(C(\Ab_\gamma))$, does not define a group homomorphism to the group of automorphisms in $C(\Ab_\gamma)$. This is only true for certain group-valued quantum flux operators, which form a flux group associated to a certain surface set. Furthermore, the analytic holonomy $C^*$-algebra can be restricted to certain subgraphs of a graph $\Gamma$. Therefore, the following object is important.
A \textit{finite orientation preserved graph system} is a set of certain subgraphs of a graph $\Gamma$ such that all paths in a subgraph are generated by compositions of the edges that generate the graph $\Gamma$. Note that in this definition the composition of edges and inverses of this edges are excluded. Then clearly there is an action of the flux group associated to the graph $\Gamma$ and a surface set on the analytic holonomy $C^*$-algebra restricted to the finite orientation preserved graph system $\PD_\Gamma^{\op}$. Furthermore, there is an action of the flux group associated to every subgraph of the finite orientation preserved graph system $\PD_\Gamma^{\op}$ and a surface set on the analytic holonomy $C^*$-algebra restricted to a finite orientation preserved graph system. There is a set of exceptional $C^*$-dynamical systems, which is defined by these automorphisms of the flux groups associated to suitable surface sets and graphs on the analytic holonomy algebras restricted to finite orientation preserved graph systems. The restriction to orientation preserved subgraphs is necessary to obtain either a purely left or right action of the flux group associated to a fixed surface set and subgraphs of a particular graph system on the holonomy $C^*$-algebra restricted to suitable graph systems. In general there are $C^*$-dynamical systems, which are constructed from left and right actions of the flux group associated to a surface set and a graph on the analytic holonomy $C^*$-algebra restricted to the finite graph system. 

The Gelfand-Na\u{\i}mark theorem implies that there is an isomorphism between commutative $C^*$-algebras and continuous function algebras on configuration spaces. If other in particular non-abelian $C^*$-algebras are studied, then automorphisms of the algebras do not correspond to certain homeomorphisms on the configuration spaces. More generally, covariant representations of the $C^*$-dynamical systems replace the construction of Fleischhack. A covariant representation is a pair of maps, which is given by a representation of the $C^*$-algebra on the Hilbert space and a unitary representation of the flux group, and these maps satisfy a certain canonical commutator relation. In this project the \textit{Weyl $C^*$-algebra for surfaces} is constructed from all $C^*$-dynamical systems, which contains all actions of the flux groups associated to all different surface sets on the analytic holonomy $C^*$-algebra. In particular an element of the \textit{Weyl algebra of a surface set $\breve S$ restricted to a finite graph system $\PD_\Gamma$} is for example of the form
\beqs &\sum_{l=1}^L\idf_\Gamma U_{S_1}(\rho^l_{S,\Gamma}(\Gamma)) + \sum_{k=1}^K\sum_{i=1}^Mf^k_{\Gamma}U_{S_2}(\rho^i_{S,\Gamma}(\Gamma))  + \sum_{k=1}^K\sum_{i=1}^MU_{S_3}(\rho^i_{S,\Gamma}(\Gamma)) f^l_{\Gamma}U_{S_3}(\rho^i_{S,\Gamma}(\Gamma))^*
+\sum_{p=1}^Pf^p_{\Gamma}
\eqs whenever $f^k_{\Gamma},f^l_{\Gamma},f^p_{\Gamma}\in C(\Ab_\Gamma)$, $U_{S_i}\in \Rep(\bar G_{\breve S,\Gamma},\KD(\HS_\Gamma))$. The notion $U_{S_i}\in \Rep(\bar G_{\breve S,\Gamma},\KD(\HS_\Gamma))$ means that the unitary operators are represented on the $C^*$-algebra $\KD(\HS_\Gamma)$ of compact operators on the Hilbert space $\HS_\Gamma$. Furthermore, the unitaries and products of these unitaries, which satisfy the canonical commutator relation, are called \textit{Weyl elements} in this project. Some further comments on the Weyl algebras and the relation to the holonomy-flux cross-product $^ *$-algebra will be given in subsection \ref{subsec Weyl}.

The use of $C^*$-dynamical systems have several advantages in comparison to the ansatz of Fleischhack, which are given by:
\begin{enumerate}
\item The operator algebraic formulation in terms of $C^*$-dynamical systems is independent of a particular Hilbert space.
\item From $C^*$-dynamical systems new algebras will be constructed. One example is constructed in \cite{Kaminski2} and is called the holonomy-flux cross-product $C^*$-algebra. Furthermore, the framwork allows to replace for example the $C^*$-algebra of quantum configuration variables.
\item The Weyl $C^*$-algebra for surfaces and the holonomy-flux cross-product $^*$-algebras will be constructed in the same framework such that the uniqueness of the state, which is invariant under certain diffeomorphisms, will be obtained easily in both cases. For a comparison of the constructions refer to \citetableA.
\item The operator algebraic framework will be used in \cite{Kaminski4} for KMS-theory in LQG (which has not been considered in the LQG framework until now).
\end{enumerate}

\subsection{The quantum spatial diffeomorphisms}\label{subsec quandiffeo}

In the project of \textit{Algebras of Quantum Variables in LQG} the classical spatial diffeomorphisms are replaced by new quantum diffeomorphism constraints. The classical diffeomorphism constraints are certain diffeomorphisms in the spatial hypersurface $\Sigma$.  In Mackenzie \cite{Mack05} a concept of translations in a general Lie groupoid has been presented. The ideas are used for a redefinition of the diffeomorphism constraints. The new operators are called bisections. The idea of the definition of a bisection is presented in the next paragraph.

In the theory of groupoids the following object is often used: the groupoid isomorphism in a path groupoid, which consists of the classical diffeomorphism in $\Sigma$ and an additional bijective map from paths to paths in the path groupoid over $\Sigma$. This pair of maps is called the \textit{path-diffeomorphisms of a path groupoid}. The path-diffeomorphisms extend the notion of  graphomorphisms, which have been introduced by Fleischhack \cite{Fleischhack06}. There is only a slight difference betweeen these objects: A graphomorphism is a map from $\Sigma$ to $\Sigma$ that preserves additionally the path groupoid structure, whereas a path-diffeomorphism is a pair of maps. In particular \textit{finite path-diffeomorphisms} are given by a pair of maps, which contains a map that maps paths to paths in a finite path groupoid $\PD_\Gamma\Sigma$ and a bijective map that maps vertices to vertices of the vertex set of the graph $\Gamma$. Moreover, since graph systems are used in this project, a pair of maps that contains a map, which maps subgraphs to subgraphs, plays a fundamental role and is called \textit{finite graph-diffeomorphism}. Graphomorphisms define in particular groupoid isomorphisms and, hence, they transform non-trivial paths to non-trivial paths. To define maps that transform a trivial path at a vertex in $\Sigma$ to a non-trivial path other objects have to be considered. Translations in a finite path groupoid are naturally given by adding or deleting edges, which generate the graph $\Gamma$. One distinguishes between three translations in a path groupoid. One is given by adding a path $\gamma$ to a path $\theta$ at the source vertex $s(\theta)$ of the path $\theta$. The other case is given by composition of a path $\gamma$ to a path $\theta$ at the target vertex $t(\theta)$ of the path $\theta$. Finally, two paths can be composed with a path at the source and target vertices simultaneously. Hence, there is a natural map from the set of vertices to the set of paths, which is called a \textit{bisection of a finite path-groupoid}. For such a bisection $\sigma$ the map $t\circ\sigma$ is assumed to be bijective, where $t$ denotes the target map of the finite path groupoid. In the definition of a \textit{bisection of a path groupoid} the map $t\circ\sigma$ is required to be a diffeomorphism from $\Sigma$ to $\Sigma$ and the map $\sigma$ maps vertices to paths in a path groupoid.  Furthermore a \textit{right-translation $R_\sigma$ of a bisection $\sigma$} is a map that composes a path $\gamma$ with the path $\sigma(t(\gamma))$, i.e. $R_\sigma(\gamma)=\gamma\circ\sigma(t(\gamma))$. Furthermore a \textit{left-translation $L_\sigma$} and an \textit{inner-translation $I_\sigma$ of a bisection $\sigma$} can be defined similarly. The pair consisiting of the composition $t\circ\sigma$ of the bisection and the target map and the right translation $R_\sigma$ define in general no groupoid isomorphism. Nevertheless there are particular translations of suitable bisections that define path-diffeomorphisms. There is no doubt that the notion of a bisection can be generalised to a \textit{bisection of a path groupoid} or a \textit{bisection of a finite graph system}.
Moreover, the bisections of a path groupoid form a group and there is a group homomorphism between this group and the group of diffeomorphisms in $\Sigma$. Moreover, the bisections of a finite path groupoid or a finite graph system equipped with a sophisticated group multiplication form groups, too. Finally, a quantum diffeomorphism is assumed to be an element of the group of bisections of a path groupoid, a finite path groupoid or a finite graph system. 

Now, actions of the group of bisections on the analytic holonomy $C^*$-algebra restricted to a finite graph system will be used in \cite[Section 3.2]{Kaminski1}, \cite[Section 6.2]{KaminskiPHD} to construct $C^*$-dynamical systems. If the group $\mathfrak{B}(\PD_\Gamma)$ of bisections of a finite graph system $\PD_\Gamma$ is considered, then the right-, left- or inner-translation of the bisections define three different $C^*$-dynamical systems. For example, there is a $C^*$-dynamical system $(C(\Ab_\Gamma),\mathfrak{B}(\PD_\Gamma),\zeta)$, where the action $\zeta$ is defined by the right-translation of the bisections. For each $C^*$-dynamical system there exists a covariant representation on the Hilbert space $\HS_\Gamma$.   Hence, the right- , left- or inner-translation of the bisections define unitary operators on the Hilbert space $\HS_\Gamma$ associated to a graph. The main advantage of translations of bisections is that they define graph changing operators. In particular these maps transfrom subgraphs into subgraphs of a graph $\Gamma$ such that the number of edges of the subgraphs can change.

Both actions, which are the action of the group of bisections of a finite graph system and the action of the flux group on the configuration space, lead to automorphisms on the analytic holonomy $C^*$-algebra. A comparison of the actions can be found in \citetableB. Similarly to actions of the flux group, the actions of the group of bisections composed with holonomy maps do not define groupoid morphisms in general. This causes no problems, since the configuration space restricted to a finite graph system $\PD_\Gamma$ is identified (naturally or in non-standard way) with $G^{\vert\Gamma\vert}$ and the right-, left- or inner-translation in the finite path groupoid transfer to \textit{right-translation $R_\sigma$, left-translation $L_\sigma$} or \textit{inner-translation $I_\sigma$ in the groupoid $G$ over $\{e_G\}$}. Finally, notice that only actions of certain bisections preserve the flux operators associated to a surface $S$. For example consider the bisection $\sigma$ of a path groupoid and recall the diffeomorphism $t\circ\sigma$. Then for example the diffeomorphism $t\circ\sigma$ is required to preserve the surface $S$. This particular bisection is called the \textit{surface-preserving bisection of a path groupoid}. There exists a similar description for a \textit{surface-preserving bisection for a finite path groupoid or a finite graph system}. Then the concept can be extended to bisections of a finite graph system that map surfaces to surfaces in a certain surface set and preserve the orientation of the surfaces with respect to the transformed subgraph. In this situation the bisections are called \textit{surface-orientation-preserving bisections for a finite graph system} and they form a subgroup of the group of bisection of a finite graph system.
Finally both actions on the analytic holonomy $C^*$-algebra restricted to a finite graph system:
\begin{enumerate}
\item the action of the group of surface-orientation-preserving bisections for a finite graph system and
\item the action of the center of the flux group associated to a surface set 
\end{enumerate} commute. In analogy to the surface-orientation-preserving bisections of a finite graph system the \textit{surface-orientation-preserving graph-diffeomorphisms} can be constructed.

Finally there is an action of bisections of the path groupoid $\PD$ over $\Sigma$ or the inductive limit graph system $\PD_{\Gamma_\infty}$ on the analytic holonomy $C^*$-algebra $C(\Ab)$. This automorphism is not point-norm continuous. Consequently, the infinitesimal diffeomorphism constraint is not implemented as a Hilbert space operator.

\section{Algebras in Loop Quantum Gravity}\label{subsec algLQG}

\subsection{The Weyl $C^*$-algebras and the holonomy-flux $^*$-algebras}\label{subsec Weyl}

The main objects, which are introduced in this project, are given by
\begin{itemize}
 \item the flux groups  or the Lie flux algebras of Lie flux groups associated to surface sets,
 \item the analytic holonomy $C^ *$-algebra, which is given by the inductive limit $C^ *$-algebra of an inductive family of analytic holonomy $C^ *$-algebras restricted to finite graph systems.
\end{itemize} 
They are used for the definition of the Weyl $C^ *$-algebra for surfaces, the holonomy-flux cross-product $C^ *$-algebra and the holonomy-flux cross-product $^ *$-algebra. These three algebras are constructed by using different representations of the flux group, or of functions depending on elements of the flux group or of the Lie flux algebra and the representation of the analytic holonomy $C^ *$-algebra in the $C^ *$-algebra $\LD(\HS_\infty)$ of bounded operators on the inductive limit Hilbert space $\HS_\infty$. The ideas for the development of the Weyl $C^ *$-algebra for surfaces and the holonomy-flux cross-product $^ *$-algebra are presented in the next paragraphs. A detailed overview about the correspondence between the two algebras will be presented in \citetableC. The following degrees of freedom will be used for a construction of an algebra in LQG: \ref{listofoptions1}, \ref{listofoptions5} and \ref{listofoptions7}.

In subsection \ref{subsec fluxop} the construction of the Weyl algebra was introduced. The Weyl algebra of Quantum Geometry \cite{Fleischhack06} has been constructed from the analytic holonomy $C^*$-algebra and unitary operators, which are defined by weakly continuous one-parameter unitary groups of $\R$ on the Hilbert space $\HS_{\text{AL}}$. The unitaries have been called Weyl operators by Fleischhack. 
The Weyl $C^*$-algebra for the surface set and restricted to a finite graph system is generated by the analytic holonomy $C^*$-algebra restricted to a finite graph system and Weyl elements. Assume that $G$ is a compact connected Lie group. Then consider a strongly continuous one-parameter unitary group of $\R$, which is given by $\R\ni t\mapsto U(\exp(t E_S(\gamma))$, on the Hilbert space $\HS_\infty$. Then each unitary $U(\exp(t E_S(\gamma))$ defines a \textit{Weyl element}, too. 

To obtain a uniqueness result of a representation of a $C^*$-algebra the following general facts will be used.
Since irreducible representations of a $C^*$-algebra on a Hilbert space correspond one-to-one to pure states on the $C^*$-algebra, the uniqueness of a particular representation of the $C^*$-algebra on a Hilbert space corresponds to a unique state. The inductive limit of an inductive family of $C^*$-algebras corresponds one-to-one to a projective limit on the projective family of state spaces of the $C^*$-algebras. The GNS-representation associated to a state of a $C^ *$-algebra consists of a cyclic vector $\Omega$ on a Hilbert space and a representation of the $C^*$-algebra on the Hilbert space. 

The uniqueness of a finite surface-orientation-preserving graph-diffeomorphism invariant pure state of the commutative Weyl $C^*$-algebra for surfaces is obtained in \cite[Theorem 3.63]{Kaminski1},\cite[\reftheouniquweylalg]{KaminskiPHD} by several steps. The \textit{commutative Weyl $C^*$-algebra for surfaces} is constructed similarly to the Weyl $C^*$-algebra for surfaces with the difference that the group $G$ is replaced by the center of the group $G$. Then graph-diffeomorphism invariant states of the commutative Weyl algebra for surfaces restricted to a graph system $\PD_\Gamma$ are analysed. It turns out that a difference occur, if either the natural or if the non-standard identification of the configuration space $\Ab_\Gamma$ is taken into account. In particular, for the natural identification one state is a sum over states, which are indexed by bisections. For the commutative Weyl algebra for surfaces the difference disappears. There exists a pure and unique state, which is invariant under finite graph-diffeomorphisms. This result is similar to the uniqueness of the representation of the Weyl algebra of Quantum Geometry and it is obtained in a complete new operator algebraic formulation. 

Furthermore, a comparable uniqueness result of the holonomy-flux cross-product $^*$-algebra is achieved in the operator algebraic framework, too. The uniqueness is directly related to the uniqueness result of the Weyl algebra for surfaces. The holonomy-flux cross-product $^*$-algebra \cite{Kaminski3}, \cite[Section 8.2]{KaminskiPHD} is related to the holonomy-flux $^*$-algebra \cite{LOST06}. This new $^*$-algebra is, in particular, an abstract cross-product algebra. This mathematical object has been presented by Schm\"udgen and Klimyk \cite{KlimSchmued94} in the context of Hopf algebras. Similarly to the $^*$-algebras in Quantum Mechanics, which were presented in subsection \ref{subsec QM}, the new \textit{holonomy-flux cross-product $^*$-algebra} is generated by the identity $\idf$, the holonomies along paths and the Lie algebra-valued quantum flux operators satisfying the canonical commutator relations \eqref{eq fluxcross}. If the surfaces are restricted to a certain set of surfaces, then this algebra is called the \textit{holonomy-flux cross-product $^*$-algebra associated to a surface set}. In contrast to the holonomy-flux $^*$-algebra the construction of the holonomy-flux cross-product $^*$-algebra is independent of the Hilbert space and the representation of the operators on the Hilbert space. For the definition of the \textit{holonomy-flux cross-product $^*$-algebra for a graph $\Gamma$ and a surface set $\breve S$} the enveloping flux algebra $\bar\E_{\breve S,\Gamma}$ associated to a surface set $\breve S$ and a graph $\Gamma$ is necessary. Then this abstract cross-product $^*$-algebra is given by the tensor vector space of the analytic holonomy $C^*$-algebra restricted to a graph and the enveloping flux algebra associated to a surface set equipped with a multiplication operation, which is derived from a certain action of enveloping flux algebra associated to a surface set on the analytic holonomy $C^*$-algebra restricted to a graph. In particular, it is used that the analytic holonomy $C^*$-algebra restricted to a graph is a right (or left) $\bar\E_{\breve S,\Gamma}$-module algebra. 

The $^*$-representation of the enveloping flux algebra associated to a surface set and a graph is given by \textit{infinitesimal representation of the flux group associated to the surface set $\breve S$ and the graph $\Gamma$} on the Hilbert space $\HS_\Gamma$. The $^*$-representation $\pi$ of the holonomy-flux cross-product $^*$-algebra associated to the graph $\Gamma$ and the surface set $\breve S$ is given by this representation $\dif U_{\overrightarrow{L}}$ of the enveloping flux algebra associated to the surface set and the graph and the representation $\Phi_M$ of the analytic holonomy $C^*$-algebra restricted to the graph. Consequently, an element $f_\Gamma\otimes E_S(\gamma)$ is represented on the Hilbert space $\HS_\Gamma$ by
\beqs \pi(f_\Gamma\otimes E_S(\gamma)):=\frac{1}{2}\Phi_M(e^{\overrightarrow{L}}(f_\Gamma)) + \frac{1}{2} \Phi_M(f_\Gamma)\dif U_{\overrightarrow{L}}(E_S(\gamma))
\eqs
where $e^{\overrightarrow{L}}$ denotes the right-invariant vector field and $E_S(\gamma)$ is an element of the enveloping flux algebra associated to a surface set $\breve S$ and a graph $\Gamma$. The representation extends to a representation of the  holonomy-flux cross-product $^*$-algebra associated to a surface set $\breve S$. In \cite[Theorem 4.8]{Kaminski3}, \cite[Theorem 8.2.20]{KaminskiPHD} it is shown that the corresponding state is the unique surface-orientation-preserving graph-diffeomorphism invariant state of the holonomy-flux cross-product $^*$-algebra associated to the surface set $\breve S$.

In particular, the analysis shows the reason for the difficulty of a construction of other representations of this $^*$-algebra. One searches for other $^*$-representation of the enveloping flux algebra associated to a surface set and a graph, which satisfy a certain graph-diffeomorphism invariance condition and which are distinguished from an infinitesimal representation. In particular, since the right-invariant vector fields associated to a surface set and a graph define a $^*$-derivation $\delta$ on the analytic holonomy $C^*$-algebra restricted to a graph, the corresponding state $\omega$ of the representation of the analytic holonomy $C^*$-algebra is assumed to satisfy $\omega(\delta(f_\Gamma))\neq 0$ for every $f_\Gamma\in C^\infty(\Ab_\Gamma)$. Consequently the state associated to the $^*$-representation of the holonomy-flux cross-product $^*$-algebra restricted to the analytic holonomy $^*$-algebra is required to satisfy a similar condition. In the project \textit{AQV} the conditions for such states associated to new $^*$-representations are presented, but the states, or respectively the representations, are not explicitly constructed.

The same problem of finding other representations of the algebras occurs for the Weyl $C^*$-algebra for surfaces or the holonomy-flux cross-product $C^*$-algebra, too. For example for the Weyl $C^*$-algebra for surfaces the important fact is that the flux group associated to a surface set is represented on the Hilbert space $\HS_\Gamma$ by a unitary representation in $\Rep(\bar G_{\breve S,\Gamma}, \KD(\HS_\Gamma))$. The only naturally or satisfactory representations of the group- (or enveloping algebra-)valued quantum flux operators are given by:
\begin{enumerate}
 \item Weyl elements, which are given by unitary representation of the flux group on the Hilbert space $\HS_\Gamma$,
\item the differential operators, which are given by the infinitesimal representation of the enveloping flux algebra on the Hilbert space $\HS_\Gamma$.
\end{enumerate} These representations define the natural representations of the following algebras:
\begin{enumerate}
 \item the Weyl $C^*$-algebra for surfaces,
 \item the holonomy-flux cross-product $^*$-algebra.
\end{enumerate} 

Finally remark that, the problem of finding other representations can be solved if the generating set of operators for the algebras does not contain unitary or differential operators derived from the flux group associated to a surface set. A new idea for a solution has been introduced by Buchholz and Grundling in \cite{BuchholzGrundling07}. They have proposed a resolvent $C^*$-algebra in the context of QFT. This $C^*$-algebra is generated by the resolvents of the Segal operators instead of unitaries, which generate the original Weyl $C^*$-algebra. In this project a similar construction of $C^*$-algebras generated by a set of operators and relations among them will be briefly presented in the next section. 

Finally, the same objects, which define the Weyl algebra for surfaces, generate the \textit{holonomy-flux von Neumann algebra} \cite{Kaminski4}, \cite[Section 6.5]{KaminskiPHD}. Note that this feature is related to the degree of freedom \ref{listofoptions3}.

\subsection{The holonomy-flux cross-product $C^*$-algebras, other cross-product $C^*$-algebras and other $^*$- or $C^*$-algebras}\label{subsec holfluxcrossprodCalg}

\paragraph*{The cross-product $C^*$-algebras for holonomies, fluxes and graph-diffeomorphisms\\}\hspace{10pt}

There is no obvious reason why the Weyl algebra of Quantum Geometry or the Weyl $C^*$-algebra for surfaces are the exceptional $C^*$-algebras of quantum configuration and momentum operators in LQG. New $C^*$-algebras of quantum variables in LQG are given by the holonomy-flux cross-product $C^*$-algebra for a surface set and the multiplier algebra of the holonomy-flux cross-product $C^*$-algebra for a surface set. These algebras will be introduced briefly in the next paragraphs and will be constructed in \cite[Chapter 2]{Kaminski2}, \cite[Section 7.2]{KaminskiPHD}. For a comparison of the Weyl $C^*$-algebra for surfaces and the new holonomy-flux cross-product $C^*$-algebra refer to \citetableD. In particular the new algebras will be defined by using the degrees of freedom \ref{listofoptions3}, \ref{listofoptions4} and \ref{listofoptions5}. 

Recall for a moment the construction of the Weyl $C^*$-algebra for surfaces of the last section. There the requirement of the group-valued flux operators to be unitary Hilbert space operators was the important starting point. If this choice is not made, then the group-valued quantum flux operators can be represented on the Hilbert space $\HS_\Gamma$ by the \textit{generalised group-valued flux operators}, which are given by the integrated representations of the flux group associated to a surface set and the graph $\Gamma$ on the Hilbert space $\HS_\Gamma$. Furthermore, consider instead of the group-valued quantum flux operators contained in the flux group $\bar G_{\breve S,\Gamma}$ for a surface set and a graph, certain functions which depend on the flux group and which map to the algebra $C(\Ab_\Gamma)$. These functions form the $^*$-algebra $L^1(\bar G_{\breve S,\Gamma}, C(\Ab_\Gamma))$. This $^*$-algebra equipped with the $L^1$-norm is, in particluar, a Banach $^*$-algebra. Then a representation of the Banach $^*$-algebra $L^1(\bar G_{\breve S,\Gamma}, C(\Ab_\Gamma))$ on the Hilbert space $\HS_\Gamma$ is derived from the unitary representations $\Rep(\bar G_{\breve S,\Gamma}, \KD(\HS_\Gamma))$ and is called the \textit{Weyl-integrated holonomy-flux representation} on $\HS_\Gamma$. This new representation is used for the definition of the \textit{holonomy-flux cross-product $C^*$-algebra associated to the surface set $\breve S$ and the graph $\Gamma$}. The construction of this algebra uses a particluar action of a fixed flux group $\bar G_{\breve S,\Gamma}$ on the analytic holonomy $C^*$-algebra $C(\Ab_\Gamma)$ restricted to a finite graph system $\PD_\Gamma$, and hence a particular $C^*$-dynamical system and depends highly on the fixed choice of the surface set $\breve S$. If another surface set $\breve T$ is considered, then a new holonomy-flux cross-product $C^*$-algebra associated to the surface set $\breve T$ and the graph $\Gamma$ can be constructed. An element of this algebra is not contained in general in the holonomy-flux cross-product $C^*$-algebra associated to the surface set $\breve S$ and the graph $\Gamma$.

Indeed there are many distinguished $C^*$-dynamical systems for different surface sets (refer to \cite[Section 3.1]{Kaminski1}, \cite[Section 6.1]{KaminskiPHD}) and consequently there exists a set of holonomy-flux cross-product $C^*$-algebras associated to different surface sets. A particular surface set $\breve S$ is chosen such that $\breve S$ has the \textit{simple surface intersection property for a graph $\Gamma$}. This means that each path in $\Gamma$ intersects only one surface in $\breve S$ only once in the target vertex of the path. There are no other intersection points between paths and surfaces. Then it will be proved in \cite[Theorem 4.9 with an extended proof]{Kaminski3}, \cite[\refpropmultilpiercrossprod  with main arguments for the proof given in Remark 7.2.10]{KaminskiPHD} that the \textit{multiplier algebra of the holonomy-flux cross-product $C^*$-algebra associated to the surface set $\breve S$ and the graph $\Gamma$} contains all operators of the holonomy-flux cross-product $C^*$-algebras associated to other suitable surface sets and the graph $\Gamma$. The idea of the proof is the following.

An element of this multiplier algebra is a linear map from the holonomy-flux cross-product $C^*$-algebra associated to a surface set $\breve S$ and the graph to the holonomy-flux cross-product $C^*$-algebra associated to a surface set $\breve S$ and the graph that satisfies a certain condition, which is connected to the existence of an adjoint operator. Hence, one has to show that particular maps are multipliers. Such linear maps can be for example given by the (left) multiplication of an element of the holonomy-flux cross-product $C^*$-algebra associated to a suitable surface set $\breve T$ and the graph. Note that $\breve T$ can be chosen to be equal to $\breve S$. It is clear that if the multiplier algebra of another holonomy-flux cross-product $C^*$-algebra associated to a surface set $\breve R$ and the graph is considered, then an element of the holonomy-flux cross-product $C^*$-algebra associated to the surface set $\breve S$ and the graph can be an element of this multiplier algebra, too. But an element of the multiplier algebra of the holonomy-flux cross-product $C^*$-algebra associated to the surface set $\breve S$ and the graph is in general not an element of the multiplier algebra of the holonomy-flux cross-product $C^*$-algebra associated to the surface set $\breve R$ and the graph. 

In \cite[Section 4]{Kaminski2}, \cite[Section 7.2]{KaminskiPHD} states on a holonomy-flux cross-product $C^*$-algebra that depend on the choice of the surface set are presented. Hence they are not generally path- or graph-diffeomorphism invariant. 

Furthermore, assume $G$ to be compact. Then there exists an inductive family of holonomy-flux cross-product $C^*$-algebras associated to the fixed surface set $\breve S$ and graphs, which defines the inductive limit $C^*$-algebra. This $C^*$-algebra is called the \textit{holonomy-flux cross-product $C^*$-algebra for the surface set $\breve S$}. This $C^*$-algebra is in particular constructed from the analytic holonomy $C^*$-algebra, the flux group associated to the fixed surface set and the particluar action of this group on the analytic holonomy $C^*$-algebra. Similarly to the multiplier algebra of a cross-product C$^*$-algebra associated to the surface set and a fixed graph, the \textit{multiplier algebra of the holonomy-flux cross-product $C^*$-algebra for the surface set $\breve S$} contains all elements of the holonomy-flux cross-product $C^*$-algebras for other suitable surface sets. Moreover, the multiplier algebra of the holonomy-flux cross-product $C^*$-algebra for the suitable fixed surface set contains the holonomy-flux cross-product $C^*$-algebra for the surface set and the holonomy-flux cross-product $C^*$-algebra for the surface set and a graph.

In the last paragraphs new $C^*$-algebras of a special kind have been constructed. All these algebras are based on new operators, which are more general than group-valued quantum flux operators and which take, in particluar, values in the analytic holonomy $C^*$-algebra. Until now the quantum diffeomorphisms are implemented only as automorphisms on these algebras. In the following paragraphs one of the previous algebras is extended such that functions on the group of bisections of a finite graph system to the holonomy-flux cross-product $C^*$-algebra, form this new $C^*$-algebra.

The cross-product $C^*$-algebra construction is particularly based on $C^*$-dynamical systems. In subsection \ref{subsec quandiffeo} it was argued that the action of the group of bisections of a finite graph system on the analytic holonomy $C^*$-algebra restricted to a finite graph system define a $C^*$-dynamical system, too. Furthermore, there is also an action of the group of certain bisections of a finite graph system on the holonomy-flux cross-product $C^*$-algebra associated to the surface set $\breve S$ and a graph. These objects define another $C^*$-dynamical system and a new cross-product $C^*$-algebra, which is called the \textit{holonomy-flux-graph-diffeomorphism cross-product $C^*$-algebra} in \cite[Section 7.3]{KaminskiPHD},\cite{Kaminski2}.

There exists a covariant representation of this $C^*$-dynamical system on a Hilbert space. This pair is given by a unitary representation of the group of surface-orientation-preserving bisections of a finite graph system on the Hilbert space $\HS_\Gamma$ and the multiplication representation $\Phi_M$ of the analytic holonomy $C^*$-algebra restricted to the finite graph system $\PD_\Gamma$ on $\HS_\Gamma$. The unitaries are called the \textit{unitary bisections of a finite graph system and surfaces} in the project \textit{AQV}. Then each unitary bisections of a finite graph system and surfaces is contained in the \textit{multiplier algebra of the holonomy-flux-graph-diffeomorphism cross-product $C^*$-algebra associated to a graph and the surface set}. The remarkable point is that the multiplier algebra of the holonomy-flux cross-product $C^*$-algebra associated to a graph and the surface set does not contain these unitaries. 

In general, the multiplier algebra of the holonomy-flux cross-product $C^*$-algebra associated to a fixed surface set contains all operators of holonomy-flux cross-product $C^*$-algebra for other suitable surface sets, elements of the analytic holonomy $C^*$-algebra and all Weyl elements associated to other suitable surface sets. The Weyl $C^*$-algebra for surfaces contains elements of the analytic holonomy $C^*$-algebra and all Weyl elements. The multiplier algebra of the holonomy-flux cross-product $C^*$-algebra associated to the surface set $\breve S$ contains the Weyl algebra for suitable surface sets. The Lie algebra-valued quantum flux operators and the right-invariant vector fields are affiliated with the holonomy-flux cross-product $C^*$-algebra, but they are not affiliated with the Weyl $C^*$-algebra for surfaces. For a detailed overview about the multiplier algebras and affiliated elements with the $C^*$-algebras of quantum variables refer to \citetableF.\\

\paragraph*{The cross-product $C^*$-algebras for holonomies or fluxes\\}\hspace{10pt}

The cross-product $C^*$-algebra construction depends on the choice of the quantum configuration and momentum variables. The quantum configuration and momentum variables and the algebras will be studied separately from each other in the next paragraphs and will be studied explictly in \cite[Section 3]{Kaminski2}, \cite[Section 7.1]{KaminskiPHD}.

First consider only the group-valued quantum flux operators that define a flux group associated to a graph and a surface set. Then there exists a certain cross-product $C^*$-algebra, which is only derived from quantum flux operators and which is, therefore, called the \textit{flux transformation group $C^*$-algebra associated to a graph and a surface set}. 

A cross-product $C^*$-algebra derived only from holonomies along paths of a graph is called the \textit{heat-kernel-holonomy $C^*$-algebra}. The name of this algebra is influenced by the work of Ashtekar and Lewandowski \cite[section 6.2]{AshLewDG95}, where the authors have studied heat kernels.  This algebra is distinguished from the analytic holonomy $C^*$-algebra. The heat-kernel-holonomy $C^*$-algebra associated to a graph contains certain functions on the configuration space $\Ab_\Gamma$ to the analytic holonomy $C^*$-algebra.

All algebras defined in the previous paragraph are constructed from the basic quantum variables, which are given by the holonomies along paths and the quantum fluxes. Some of them have been even indirectly proposed in LQG literature before. Hence, they are possible algebras of a quantum theory of gravity.

\paragraph*{Simplified cross-product $C^*$-algebras for holonomies and fluxes\\}\hspace{10pt}

If both quantum variables: the quantum configuration and momentum variables restricted to a fixed graph $\Gamma$ and a fixed suitable surface set $\breve S$ are considered simultaneously, then the following simplifications can be studied. 

The flux group of a fixed graph $\Gamma$ and a fixed suitable surface set $\breve S$ and the configuration space $\Ab_\Gamma$ are identified with $G^{\vert\Gamma\vert}$. Moreover, the corresponding cross-product $C^*$-algebra $C(G^{\vert\Gamma\vert})\rtimes_\alpha G^{\vert\Gamma\vert}$ is Morita equivalent to the $C^*$-algebra of compact operators on the Hilbert space $L^2(G^{\vert\Gamma\vert},\mu_\Gamma)$, where $\mu_\Gamma$ denotes the product of $\vert\Gamma\vert$ Haar measures. Therefore, the representation theory of both $C^*$-algebras is the same and, hence, there is only one irreducible representation of the cross-product $C^*$-algebra up to unitary equivalence. 

But, this identification is only true for certain surface sets. The cross-product $C^*$-algebra is derived from the quantum momentum variables, which depend on the surface sets. In particular the flux groups associated to a suitable surface set can be identified with a product group $G^M$ where $M\leq \vert\Gamma\vert$. Then there exists a left (or right) action of $G^M$ on the $C^*$-algebra $C(G^{\vert\Gamma\vert})$. For $M< \vert\Gamma\vert$ a Morita equivalent $C^*$-algebra is not found in this project. In \cite[Theorem 7.1.11]{KaminskiPHD} a Morita equivalent algebra for the $C^*$-algebra $C(G^{N})\rtimes_\alpha G^{M}$ whenever $N<M$, is given.

In this project the general case of arbitrary surfaces is studied. Hence, the quantum configuration and the momentum variables of the theory are manifestly distinguished from each other. The quantum configuration variables only depends on graphs and holonomy mappings, whereas the quantum momentum variables depend on graphs, maps from graphs to products of the structure group, and the intersection behavior of the paths of the graphs and surfaces. But, nevertheless, the elements of the holonomy-flux cross-product $C^*$-algebra will be understood as compact operators on the flux group associated to a surface set with values in the analytic holonomy $C^*$-algebra restricted to a graph, which are acting on the Hilbert space $L^2(\Ab_\Gamma,\mu_\Gamma)$. 

\paragraph*{Other $^*$- or $C^*$-algebras for holonomies, fluxes and other flux operators\\}\hspace{10pt}

There are two different $^*$-algebras, which can be completed to $C^*$-algebras, which contains certain continuous functions on a locally compact group. The difference between the $^*$-algebras
are related to the choice of the pointwise multiplication 
or the convolution multiplication operation. These two different $^*$-algebras can be completed to two different $C^*$-algebras. In the project \textit{AQV} the analytic holonomy $C^*$-algebra is obtained from the pointwise multiplication and the \textit{non-commutative holonomy $C^*$-algebra} is obtained from the convolution operation. Note that this is related to the degrees of freedom \ref{listofoptions2} and \ref{listofoptions3}. For a compact group this issue will be studied in \cite[\refsecalgperiod]{KaminskiPHD} explicitly. In Quantum Mechanics the locally compact group is replaced by the abelian locally compact group $\R^n$. Then these two $C^*$-algebras are isomorphic (refer to \cite[\refsecAPLQC]{KaminskiPHD}). The two $C^*$-algebras obtained by the two different multiplication operations are not isomorphic for arbitrary non-abelian locally compact groups. Hence, in general in the project \textit{AQV} the analytic holonomy $^*$-algebra and the non-commutative holonomy $^*$-algebra are not isomorphic. 

In the LQG literature, the compact group $SU(2)$ has been often used, but it is not the only structure group, which has been studied. For example, the non-compact group $\text{SL}(2,\CB)$ in \cite{AshIsh92} or the compact quantum group $SU_q(2)$ in \cite{LewOk08} have been used, too. In particular, Lewandowski and Oko\l{}\'{o}w \cite{LewOk08} have used the non-commutative holonomy $C^*$-algebra, which they construct from the quantum group. Hence in the LQG framework, the difficulties, which arise by replacing the compact connected Lie group by other groups or quatum groups, have to be analysed. This is the reason for the choice of a locally compact structure group $G$ in the project \textit{AQV} for the construction of the Weyl algebras for surfaces and the holonomy-flux cross-product $C^*$-algebras. Clearly, for more general groups and in particular for $\text{SL}(2,\CB)$ the development has to be studied in detail once more. The starting point of the construction of the holonomy-flux cross-product $^*$-algebra and other $^*$-algebras is a compact Lie group.

Furthermore, new ideas for a construction of $C^*$-algebras are available by the concept of affiliated operators. For example Woronowicz \cite{Woro91} has developed a construction of $C^*$-algebras by a finite set of bounded or unbounded operators. Moreover, Buchholz and Grundling \cite{BuchholzGrundling07} have introduced a new $C^*$-algebra in the context of QFT. These ideas will be used for the definition of new algebras, which will be presented briefly in the next paragraphs.

On the one hand, the Lie algebra-valued quantum flux operators for a surface in a surface set and a fixed graph are replaced once more by new operators. The new operators are given by the \textit{flux Nelson transforms for a surface set and a fixed graph}, which are similarly to the resolvents of Buchholz and Grundling or the $Z$-transforms of Schm\"udgen or Woronowicz. The \textit{holonomy-flux Nelson transform $C^*$-algebra associated to a surface set and a fixed graph} in \cite[Section 8.5]{KaminskiPHD} is generated by the operators: holonomies along paths of a fixed graph and the flux Nelson transform for a surface set and a fixed graph and canonical commutator relations similarly to \eqref{eq fluxcross}.

On the other hand instead of the Lie algebra-valued quantum flux operators for a surface in a surface set and a fixed graph, the functions depending on holonomies along paths are replaced by polynomials of, or representative functions depending on, holonomies along paths of a fixed graph. Then for example a new $^*$-algebra can be constructed by the operators: polynomials of holonomies along paths of a fixed graph and Lie algebra-valued quantum flux operators for a surface in a surface set and the fixed graph and canonical commutator relations similarly to \eqref{eq ccr}. This new $^*$-algebra is called the \textit{Heisenberg polynomial-holonomy-flux $^*$-algebra associated to a graph and a surface set} in \cite[Section 8.2.2]{KaminskiPHD}, \cite[Section 3.2]{Kaminski3}. The name of the algebra is influenced by Schm\"udgen and Inoue, who have defined the Heisenberg $O^*$-algebra in Quantum Mechanics for example in \cite{Inoue}. The important degree of freedom for this construction is \ref{listofoptions2}. If all continuous functions in $C^\infty(\Ab_\Gamma)$ are considered, then it is possible to construct the \textit{Heisenberg holonomy-flux $^*$-algebra associated to a graph and a surface set}, which contains these functions and the quantum fluxes associated to a surface set $\breve S$ and a graph $\Gamma$ with values in the enveloping algebra $\bar\E_{\breve S,\Gamma}$ and which satisfy some canonical commutator relation. This relation is distinguished from the canonical commutator relation of the holonomy-flux $^*$-algebra associated to a graph and a surface set.

Notice that these algebras are derived from the basic quantum variables of LQG and are completely new in this framework. 

\paragraph*{The important question\\}\hspace{10pt}

Finally, the important question is the following: Which algebra is the algebra of a quantum theory of gravity? Hence, physical reasons have to be taken into account to answer the question why some algebras are more suitable than others. 

\subsection{The quantum constraints of Loop Quantum Gravity}\label{subsec quantumconstraints}

In the LQG framework the quantum constraint algebra is generated by the quantum gauge constraints, the quantum diffeomorphism constraints and the quantum Hamilton constraint.
The quantum gauge constraints are replaced by elements of the \textit{local flux group}. The elements of the \textit{fixed point algebra associated to the action of the local flux group} \cite[Section 6.2]{KaminskiPHD} are invariant under the action of the local flux group. In the next subsection the algebra, which is generated by the quantum diffeomorphism constraints, is analysed.

\paragraph*{The quantum spatial diffeomorphism constraints\\}\hspace{10pt}

In subsection \ref{subsec quandiffeo} the quantum spatial diffeomorphism constraints have been introduced as bisections of a path groupoid, a finite path groupoid or a finite graph system. It has been argued that, the group $\Diff(\Sigma)$ of classical diffeomorphisms in the spatial manifold $\Sigma$ is replaced by the group $\mathfrak{B}(\PD)$ of bisections in a path groupoid $\PD$ over $\Sigma$. In the next paragraph this issue is treated once more.

There exists a parameter group, which is given by $\Diff(\Sigma)\ni \varphi\mapsto \zeta_\varphi\in\Aut(\Alg)$, of automorphisms on the commutative Weyl $C^*$-algebra $\Alg$ for surfaces of Loop Quantum Gravity. Indeed, a diffeomorphism $\varphi$ in $\Diff(\Sigma)$ is given by a bisection $\sigma$ in $\mathfrak{B}(\PD)$ through the map $t\circ\sigma$. Consequently, this parameter group of automorphism is reformulated by the parameter group $\mathfrak{B}(\PD)\ni\sigma\mapsto\zeta_\sigma\in\Aut(\Alg)$ of automorphism on $\Alg$. The remarkable property of this parameter group is that, the automorphism $\zeta$ of the $C^*$-algebra $\Alg$ is not point-norm continuous. This is verified by the following argument: For a sequence of paths that converges to the constant path at a vertex, the sequence of holonomy maps along the paths does not converge. Hence, finite classical diffeomorphisms and bisections of a finite path groupoid have to be considered. The parameter group of automorphism defined by morphisms from the group of bisections on a finite path groupoid to the automorphism group on the $C^*$-algebra $\Alg$, is point-norm continuous.  Note that, since the Weyl algebra for surfaces is constructed from an inductive limit of an inductive family of $C^*$-algebras restricted to finite graph systems, the group of bisections are related to finite graph systems instead of finite path groupoids.

In subsection \ref{subsec holfluxcrossprodCalg} it has been discussed that, the unitary bisections of a finite graph system and surfaces are neither contained in the Weyl $C^*$-algebra for surfaces nor in the holonomy-flux cross-product $C^*$-algebra. Similarly, the finite surface-orientation-preserving graph-diffeomorphisms, or respectively the surface-orientation-preserving bisections of a finite graph system, are not contained in these algebras. 
But they are affiliated with a larger $C^*$-algebra, which is given by the holonomy-flux-graph-diffeomorphism cross-product $C^*$-algebra \cite[\refsubsecholfluxdiffcrossalg]{KaminskiPHD}. 

\paragraph*{The quantum Hamilton constraint\\}\hspace{10pt}

In subsection \ref{subsubsec MasterHamiltonconstr} the classical Hamilton constraint $C(x)$ of LQG has been shortly introduced. In the next paragraphs the quantisation of this constraint, which has been proposed by Thiemann \cite{Thiem96}, is reviewed. 

Thiemann has presented the following classical expression for the classical Hamilton constraint
\beq\label{eq Hamconstraint} C(x)=\frac{1}{\sqrt{det(q)}}tr(\left(F_{ab} -[K_a,K_b]\right)[E^a,E^b])\eq
where $F_{ab}$ is the curvature of the connection $A$, $\beta K_a= A_a-\Gamma_a$ is the extrinsic curvature and $\beta$ is the Immirzi parameter.

Then Thiemann has used the classical identity:
\beqs \frac{[E^a,E^b]}{\sqrt{det(q)}}=\epsilon^{abc}\{A_c,V\}
\text{ with } V=\int \dif^3x \sqrt{det(q)}\eqs The volume $V$ is encoded in the quantum operator $\QD(V)$, which is mainly given by a product of fluxes.

Then the quantum map of the Poisson bracket $\{A,V\}$ is given by
\beqs \QD(\{A,V\})=\ho_A(e_\Delta)[\ho_A(e_\Delta)^{-1},\QD(V)]\eqs
where $\ho_A(e_\Delta)$ denotes the holonomy of a connection $A$ along a path $e_\Delta$, which is given by the triangulation $T$ of $\Sigma$.
Moreover, the quantum operator associated to the classical variable $F$ is presented by
\beqs \QD(2\epsilon^2 F(x))= \ho_{A}(l_\Delta)-\ho_{A}(l_\Delta)^{-1} \eqs
where $l_\Delta$ is a loop at some base point $x$ and which is given by the triangulation $T$ of $\Sigma$.

The quantum Hamilton constraint for a suitable triangulation $T$ of $\Sigma$ is given by
\beq\label{eq QuantumHamiltonTriang} \QD(C(N))_T=\sum_{\Delta\in T}Tr(\left(\ho_A(l_\Delta)-\ho_A(l_\Delta)^{-1}\right)\ho_{A}(e_\Delta)[\ho_{A}(e_\Delta)^{-1},\QD(V)])\eq
The triangulation $T$ of a manifold $\Sigma$ and the choice of loops $l_\Delta$ and edges $e_\Delta$ with respect to the triangulation define a graph $\Gamma$. Finally, the quantum Hamilton constraint operator with respect to the whole hypersurface $\Sigma$ is presented by a (norm-)limit of operators
\beq\label{eq QuantumHamilton} \QD(C(N)):=\lim_{T\rightarrow \Sigma}\QD(C(N))_T\eq 

The quantum Hamilton constraint operator and some other Hamiltonians have been analysed by Thiemann in many articles. For example the Hamilton constraint has been implemented in \cite{Thiem96,Thiemann98,Thiemann97,Thiemann1997,ThiemannQSDIV1997,Thiemann97QSDVI,ThiemannMay2000} by using a certain regularisation of the constraint operator and particular triangulations of the manifold. Furthermore, Aastrup, Grimstrup and Nest \cite{AastGrimNest08,AastGrimNest08_I,AastGrimNest2008_II,AastrupGrimstrup2009} have pointed out that the holonomies along paths are differentiable continuous function on $\Ab$ with values in a finite matrix algebra $M_n(\CB)$ and hence the quantum Hamilton constraint  is a (pseudo) differential operator on the algebra of differentiable continuous function on $\Ab$ with values in a finite matrix algebra $M_n(\CB)$. 
In the simpliest case the quantum Hamilton constraint operator contains the the holonomies along paths and the quantum flux operators, which are operators represented as multiplication or (pseudo) differential operators on the Ashtekar-Lewandowski Hilbert space.
If other Hilbert space representations of the operators are considered, then the existence of a well-defined quantum Hamilton constraint, which is given by a limit of the sum over all triangulations, is not clear. Similarly, the one-parameter group of automorphisms derived from the quantum Hamilton constraint on the analytic holonomy $C^*$-algebra need not be point-norm continuous. The difficulties that can arise, are investigated in the context of the localised holonomy-flux cross-product $^*$-algebra in the next section.

The question that arises is the following: If the quantum Hamilton constraint operator is well-defined, then which algebra presented in subsections \ref{subsec Weyl} or \ref{subsec holfluxcrossprodCalg}, this operator is contained in or affiliated with? The quantum Hamilton constraint operator is not contained in or affiliated with any of these algebras. Even the simpliest version of the quantum Hamilton constraint operator, which is given by 
\beq\label{eq untracedQuantumHamilton} 
\lim_{T\rightarrow \Sigma} \sum_{\Delta\in T} \left(\ho_A(l_\Delta)-\ho_A(l_\Delta)^{-1}\right)\ho_{A}(e_\Delta)[\ho_{A}(e_\Delta)^{-1},\QD(V)]
\eq is not contained in or affiliated with any of the algebras presented before. In the next section new algebras are developed. Then a certain modificated quantum Hamilton constraint operator is related to a new $^*$-algebra, which is called the localised holonomy-flux cross-product $^*$-algebra, and which is introduced in the next section.

\subsection{KMS-Theory and a physical algebra for Loop Quantum Gravity}\label{subsec KMStheory}
\paragraph*{The physical algebra of quantum variables for LQG\\}\hspace{10pt}

In particluar in Loop Quantum Gravity the theory of KMS-states is inseparable from the problem of time evolution, the implementation of the quantum constraints and the issue of the physical algebra. Briefly the\textit{ set of conditions for a physical algebra of quantum variables} is assumed to be given by 
\begin{enumerate}
\item the quantum constraint operators are affiliated with or contained in the physical algebra and
\item the physical algebra contains complete quantum observables.
\end{enumerate}
In particular, complete quantum observables are derived from Dirac states and Dirac observables, which are defined by the constraint operators. Furthermore, KMS-states, states that define time averages, or states that define expectations of the time of occurence of an event of the physical algebra of quantum variables have to be studied.

In the next paragraphs the issue of KMS-states of the algebras presented in subsections \ref{subsec Weyl} or \ref{subsec holfluxcrossprodCalg} is analysed. In the mathematical theory of KMS-states modular objects play a fundamental role. These objects are given by the modular automorphism group and the modular conjugation. Hence, in particular modular automorphisms of the Weyl $C^*$-algebra for surfaces are studied.

\paragraph*{KMS-Theory for algebras of quantum variables for LQG\\}\hspace{10pt}

The holonomy-flux von Neumann algebra \cite[Section 6.5]{KaminskiPHD} does not allow a fruitful implementation of Tomita-Takesaki theory, since for this von Neumann algebra a cyclic and separating vector is not available. For the Weyl $C^*$-algebra for surfaces, a KMS-theory can be studied for different automorphsims. The simpliest automorphism is generated by 
the exponentiated Lie algebra-valued quantum flux operator $\exp(E_S(\Gamma)^+E_S(\Gamma))$ associated to a surface $S$ and a graph, or to the limit graph $\Gamma_\infty$ of an inductive family $\{\Gamma\}$ of graphs. But, it has been shown in \cite[\reftheoweylnokmsstates]{KaminskiPHD} that, there are no KMS-states of the Weyl $C^*$-algebra for surfaces associated to this automorphism.

Since there are no other natural automorphisms on the Weyl $C^*$-algebra for surfaces, the theory of KMS-states in LQG is very hard to investigate. The non-existence of KMS-states is related to the fact that, for example on the Weyl $C^*$-algebra for surfaces the automorphism group defined by the flux operator $E_S(\Gamma)^+E_S(\Gamma)$ is inner. Since, modular automorphisms characterise the $C^*$-algebra by outer norm-continuous automorphisms, a good ansatz is to change the automorphisms. This issue is treated in the next paragraphs.

In subsection \ref{subsec quantumconstraints}, automorphisms related to finite path- or graph-diffeomorphisms have been introduced. The automorphisms on the analytic holonomy $C^*$-algebra, which is constructed from inductive families of finite graph systems, are not very sensitive on the choice of the particular graphs in the following sense. This is due to the identification of the quantum configuration space $\Ab_\Gamma$ restricted to a finite graph system $\PD_\Gamma$ with some products of the compact group $G$. Consequently, the automorphisms on the analytic holonomy $C^*$-algebra restricted to a finite graph system $\PD_\Gamma$ are maps, that map functions depending on $G^M\times \{e_G\}\times...\times \{e_G\}$-valued operators to functions depending on $G^K\times \{e_G\}\times...\times \{e_G\}$-valued operators, where $M,K\leq \vert \Gamma\vert$. There is also an automorphism on the analytic holonomy $C^*$-algebra restricted to a finite graph system $\PD_\Gamma$, which maps functions depending on $H^M\times \{e_G\}\times...\times \{e_G\}$-valued operators to functions depending on $H^K\times \{e_G\}\times...\times \{e_G\}$-valued operators, where $M,K\leq \vert \Gamma\vert$ and $H$ is a closed subgroup of $G$. But there exists no KMS-states of the analytic holonomy $C^*$-algebra restricted to a finite graph system $\PD_\Gamma$ associated to any of these automorphisms \cite[Theorem 6.5.10]{KaminskiPHD}. 

Furthermore, one can analyse Tomita-Takesaki theory for the Weyl $C^*$-algebra for surfaces. This $C^*$-algebras is constructed from inductive families of finite graph systems, too. In the previous paragraph it is argued that the modular automorphism is independent of transformations of the graph systems. Consequently, the choice of the graph system has to be such that there are suitable automorphisms, which are generated by graph-diffeomorphisms and which leave all graphs globally invariant. Then the modular automorphism should implement the dynamics of the theory, and consequently this automorphism is related to the one-parameter velocity transformation along the foliation parameter. Until now all quantum variables are implemented on a fixed Cauchy surface $\Sigma$. If the algebra of quantum variables is enlarged such that the algebra elements depend on different Cauchy surfaces, then the one-parameter group of automorphisms maps algebra elements, which depend on a certain Cauchy surface, to algebra elements, which depend on the transformed Cauchy surface. These automorphisms are suggested to map quantum configuration variables, which are defined on a fixed Cauchy surface, to quantum operators that are derived from quantum configuration and momentum variables, which are defined on the fixed Cauchy surface, too. Note that such automorphisms are not defined by either the holonomies along paths or the group-valued (or the Lie algebra-valued) quantum flux operators. These automorphisms are derived from both quantum operators.  Indeed such automorphisms should be related to the quantum Hamilton constraint or a modified quantum Hamilton constraint and are required to commute with the automorphisms related to graph-diffeomorphisms. 

But the Weyl $C^*$-algebra for surfaces does not admit a KMS-state even for the simple automorphism, which is derived from the exponentiated Lie algebra-valued quantum flux operator. Hence, the author suggests that the Weyl $C^*$-algebra for surfaces does not admit a KMS-state for an automorphism derived from the untraced version of the quantum Hamilton constraint. Consequently, the last possibility is to change the $C^*$-algebra of quantum variables or to consider $O^*$-algebras. 

Summarising, the author proposes the following ans\"atze for a KMS-theory in LQG. The analysis of the quantum constraints and their relation to the algebras of quantum variables implies that, the modular objects in Loop Quantum Gravity have to be implemented 
\begin{enumerate}
 \item\label{anaalg} on a $^*$-subalgebra of the holonomy-flux cross-product $^*$-algebra, or  
 \item on a new $^*$-algebra, which is called the localised holonomy-flux cross-product $^*$-algebra, or a new $C^*$-algebra derived from this new $^*$-algebra.
\newcounter{enumii_saved}
\setcounter{enumii_saved}{\value{enumi}}
\end{enumerate} Furthermore, 
\begin{enumerate}
 \setcounter{enumi}{\value{enumii_saved}}
 \item\label{subdiff} the group of quantum spatial diffeomorphisms has to be restricted to a subgroup of this group, and 
 \item\label{subflux} the flux group associated to a surface set has to be restricted to a closed subgroup of this flux group.
\end{enumerate}

\paragraph*{The localised holonomy-flux cross-product $^*$-algebra\\}\hspace{10pt}

The idea of the construction of the new algebra in \cite{Kaminski3} and \cite[Section 8.4]{KaminskiPHD} is influenced by the work of Thiemann and Giesel \cite{ThiemGiesel,Thiemann06,GiesThiem07,GiesTiem07IV}. They have considered the quantum Hamiltonian operator in the framework of cubic lattices and infinite $C^*$-tensor algebras. A comparison of the localised holonomy-flux cross-product $^*$-algebra and the holonomy-flux cross-product $^*$-algebra can be found in \cite[\reftablecompalg]{KaminskiPHD}. In particular in comparison with the algebras presented before the degree of freedom  \ref{listofoptions7} is used for the construction of the localised holonomy-flux cross-product $^*$-algebra.

For the definition of the localised holonomy-flux cross-product $^*$-algebra it is used that, the flux operators are manifestly localised by the surfaces in the manifold $\Sigma$. The new configuration space is divided into two parts. One of them is constructed from holonomies along paths that start or end at some given surface and is called the \textit{localised part of the configuration space}. The other part is build from holonomies along paths that do not intersect any surface in this surface set. Hence, the first configuration space is localised on surfaces, while the second is not. Furthermore, there are two different $^*$-algebras of quantum holonomy variables. One $^*$-algebra is constructed on the localised part of the configuration space and a convolution product between functions depending on this space. In particular, the $^*$-subalgebra of central functions on the localised part of the configuration space is used. The other $^*$-algebra is given by the original analytic holonomy $^*$-algebra, but is restricted to non-localised paths. These $^*$-algebras are completed to different $C^*$-algebras and the $C^*$-tensor product of these two $C^*$-algebras defines the new \textit{localised analytic holonomy $C^*$-algebra}. The $C^*$-algebra of central functions on the localised part of the configuration space is called the localised part of the localised analytic holonomy $C^*$-algebra. This certain $C^*$-algebra admits KMS-states.

There are some new flux operators, which are defined as difference operators between Lie algebra-valued quantum flux operators on different graphs, and which are called the \textit{localised and discretised flux operators associated to surfaces}. The main difference between the original Lie algebra-valued quantum flux operator defined in this project, and the localised and discretised Lie algebra-valued flux operator both restricted to a fixed graph is that the second operator is only non-trivial on paths, which are not contained in a certain subgraph. The \textit{localised enveloping flux algebra associated to a surface set} is derived from the localised and discretised flux operators. Furthermore, there exists an action of this new localised and discretised flux operator on the $C^*$-algebra of central functions on the localised part of the configuration space.

In \cite{Kaminski3}, \cite{KaminskiPHD} the theory of an abstract cross-product $^*$-algebra will be used to define a new holonomy-flux cross-product $^*$-algebra. This construction will be also used for the definition of two new localised algebras. One algebra is based on the $^*$-algebra of central functions on the localised part of the configuration space and the other is derived from the localised analytic holonomy $^*$-algebra. The abstract cross-product $^*$-algebra, which is obtained from the localised enveloping flux algebra associated to a surface set and the $^*$-algebra of central functions on the localised part of the configuration space, is called the \textit{localised part of the localised holonomy-flux cross-product $^*$-algebra}. The \textit{localised holonomy-flux cross-product $^*$-algebra} is given by the abstract cross-product $^*$-algebra, which is obtained by the the localised analytic holonomy $^*$-algebra and the localised enveloping flux algebra associated to a surface set. There are several localised holonomy-flux cross-product $^*$-algebras for different surface sets. It is also possible to construct a \textit{localised holonomy-flux cross-product $C^*$-algebra} associated to a surface set similarly to the holonomy-flux cross-product $C^*$-algebra.

In this project, the \textit{discretised quantum volume operator} $\QD(V)_{\disc}$ is constructed as a sum over Lie algebra-valued quantum flux operators indexed by a triple of paths in a graph that start at a common vertex, which is an intersection of three surfaces. 

Consider the \textit{Lie holonomy algebra}, which is constructed from the localised configuration space restricted to a graph $\Gamma$ and the non-standard identification of this space with the product group $G^{\vert\Gamma\vert}$ of a compact connected Lie group $G$. 
This Lie algebra acts on the $C^*$-algebra of central functions on the localised configuration space restricted to a graph, too. Then the $C^*$-algebra of central functions admits KMS-states with respect to this action. The \textit{modified quantum Hamilton constraint restricted to a graph} is given by
\beqs \exp(H_{\Gamma_i}^+H_{\Gamma_i}):= \left(\ho_A(\alpha)-\ho_A(\alpha)^{-1}\right)\ho_A(\gamma)[\ho_A(\gamma)^{-1},\QD(V)_{\disc}]
\eqs The \textit{modified quantum Hamilton constraint constraint} is defined in the project \textit{AQV} as the limit\\ $H:=\lim_{i\rightarrow\infty}\sum_{\Gamma_i} \exp(H_{\Gamma_i}^+H_{\Gamma_i})$ of a sum over subgraphs of a graph of the modified quantum Hamilton constraint restricted to a graph. Note that the limit graph is assumed to contain an infinite countable set of subgraphs. 

The next step is to show that this modified quantum Hamilton constraint is well-defined and is given as the generator of a strongly continuous one-parameter group of automorphisms on the localised part of the localised analytic holonomy $C^*$-algebra. The analysis of parts of the modified quantum Hamilton constraint shows that the convergence of the limit in the norm-topology is not obvious and is related to the structure of the discretised quantum volume operator $\QD(V)_{\disc}$. Summarising, the norm-convergence of the limit of $H$ is not easy to derive. The author conjectures that this limit converges and does not depend on a particular Hilbert space representation of the modified quantum Hamilton constraint. 

After the consideration of the quantum Hamilton constraint, which is given in the project \textit{AQV} by the modified quantum Hamilton constraint, a quantum Master constraint is studied in the following paragraphs.

In the previous sections translations defined by bisections of finite path groupoids or finite graph systems play a fundamental role. The most general operators, which depend on bisections of finite graph systems that preserve a surface set $\breve S_{\disc}$, are denoted by $D_{\breve S_{\disc},\Gamma}^\sigma$ and are called the \textit{localised finite quantum diffeomorphism constraints}. For example such operators can be defined similarly to elements of the holonomy-flux-graph-diffeomorphism cross-product $C^*$-algebra. The idea for these quantum constraints is to implement the complicated relations between the classical spatial diffeomorphism constraints and the classical Hamilton constraints on the quantum level.

Then the \textit{modified quantum Master constraint} is defined to be sum of the \textit{localised quantum diffeomorphism constraint}, which is given by
\beqs D_{\breve S_{\disc}}:=\limN\sum_{i=1}^{N}\sum_{\sigma_l}D^{\sigma_l,*}_{\breve S_{\disc},\Gamma_i}D^{\sigma_l}_{\breve S_{\disc},\Gamma_i}
\eqs and the modified quantum Hamilton constraint
\beqs H:=\limN\sum_{i=1}^N\exp(H_{\Gamma_i}^+H_{\Gamma_i})
\eqs This modified quantum Master constraint generalises the Master constraint, which has been studied by Thiemann \cite{Thiembook07}.

Then the following issues will be partly studied in \cite{Kaminski4} and \cite[Section 8.4]{KaminskiPHD}, and will be further completed in a new extension of the project \textit{AQV}:
\begin{itemize}
 \item the Dirac state space of the localised holonomy-flux cross-product $^*$- or $C^*$-algebra with respect to the localised quantum Master constraint,
 \item the KMS-states of the localised analytic holonomy $C^*$-algebra and the localised holonomy-flux cross-product $^*$- or $C^*$-algebra associated to the automorphism group generated by the modified quantum Hamilton constraint,
 \item the time avarage \eqref{eq timeavarage} defined by a KMS-state,
\item the \textit{localised holonomy-flux-graph-diffeomorphism cross-product $^*$-algebra for surfaces} that contains the localised finite quantum diffeomorphism constraints and all elements of the localised holonomy-flux cross-product $^*$-algebra for surfaces, and the modified quantum Hamilton constraint is affiliated (in an appropriate sense) with this algebra;
 \item the \textit{localised $^*$-algebra of complete quantum observables for surfaces}, which is derived from the localised holonomy-flux-graph-diffeomorphism cross-product $^*$-algebra for surfaces.
\end{itemize}

The issue of KMS-states can be treated, since there is a KMS-theory for $O^*$-algebras, which has been studied for example by  Inoue \cite{Inoue}. Indeed one can show that the localised holonomy-flux cross-product $^*$-algebra is an $O^*$-algebra. Until now only KMS-states for the localised part of the localised analytic holonomy $C^*$-algebra will be presented in \cite{Kaminski4}, \cite[Section 8.4]{KaminskiPHD}. The author suggests that there are also KMS-states on the localised holonomy-flux cross-product $^*$-algebra, which are similar to the KMS-states, which have been found. The first suggestion for a physical algebra is given by the localised holonomy-flux-graph-diffeomorphism cross-product $^*$-algebra for surfaces.

\paragraph*{The thermal Hamiltonian for LQG\\}\hspace{10pt}

The difficulty of the implementation of the quantum constraints is related to the fact that there is an algebra of constraints, which do not mutually commute and which are not described by a simple algebra. In particular, the quantum constraints are not contained in the algebras of quantum variables, which are usually used in LQG. Consequently, modifications  or enlargements of the algebra have to be developed for the implementation of constraints on the operator algebraic level. Moreover, for the construction the expectation of the time of occurence of an event \eqref{eq exptimeocc}, the concept of clocks that generate a new enlarged algebra, is necessary.

On the other hand, for a KMS-theory of the algebras of quantum variables, new automorphisms have to be considered. But in the previous sections it was argued that there aren't many natural candidates for automorphisms on the Weyl $C^*$-algebra for surfaces or the holonomy-flux cross-product $^*$-algebra. It is not obvious that the quantum Hamilton constraint, or respectively a modification of this operator, is the generator of the modular group associated to a KMS-state. In contrast to QFT, the thermal Hamiltonian of this theory is not a constraint of the physical system. Consequently, the Hamiltonian of a clock can be a generator of an automorphism group, too. This is related to the problem of time in Loop Quantum Gravity, since the Hamiltonian is not a true Hamiltonian, it is a constraint. The dynamics of the theory is not implemented by the Hamilton constraint, it is given by an evolution with respect to a clock. Due to the Connes cocycle theorem for von Neumann algebras, there is only one preferred time evolution and, hence, the author suggests that the thermal time of the system is related to the clock Hamiltonian instead of the quantum Hamilton constraint. Note that there is a generalised Connes cocycle theorem even for $O^*$-algebras, which has been derived by Inoue \cite{Inoue}. The thermal equilibrium state with respect to the clock is not a thermal state with respect to the automorphisms implemented by the Hamilton constraint. The thermal states with respect to clocks have to be Dirac states. But in LQG there are no obvious quantised observables in the Weyl $C^*$-algebra or the holonomy-flux von Neumann- or $^*$-algebra, which can be used as clocks. Therefore one may ask, which automorphisms on these algebras lead to self-adjoint operators and which of these operators are thermal Hamiltonians that can be physically interpreted as a clock. Note that the automorphisms will be uniquely determined up to inner automorphisms. Consequently, it is possible that there is some relation between these certain automorphisms associated to the thermal Hamiltonian of the clock and automorphisms associated to the quantum Hamilton constraint. 

But, until now, there are no natural physical clocks contained in the algebra of quantum variables. Usually matter fields are used as clocks. Since matter fields are localised objects, an idea is to study the localised algebra of complete quantum observables on surfaces. The theory, which is described by such an algebra, is not completely diffeomorphism invariant, but this invariance can be relaxed. Then only certain diffeomorphisms, which preserve the localised surfaces in which the matter fields are situated, are taken into account. Note that the surfaces, which has been studied in the context of localised algebras, are always discretised in a suitable sense. Finally the full physical algebra can be for example given by a tensor product of a matter field algebra and the localised algebra of complete quantum observables for (discretised) surfaces.

\paragraph*{A summary for a KMS-Theory and a suggestion for a physical algebra for LQG\\}\hspace{10pt}

The following important issues have been presented in the previous sections:  
\begin{itemize}
 \item a modification of the quantum configuration space,
 \item a modification of the quantum spatial diffeomorphism constraints, the quantum Hamilton and the quantum Master constraint and
 \item a suggestion for a  physical $^*$-algebra.
\end{itemize}

The new algebra is proposed to be the physical algebra of quantum gravity, if the quantum constraint Hamiltonian \eqref{eq QuantumHamilton} is taken into account. But this quantum operator is constructed from the classical Hamiltonian by several classical transformations. The original classical Hamiltonian contains the classical variables holonomy along paths, fluxes and the curvature. But until now a quantum analogue of a curvature has not been suggested. In the next section the ideas presented in \cite{Kaminski5,KaminskiPHD} for a construction of an algebra, which is generated by curvature, connections and fluxes for a given principal fibre bundle, will be presented.

\subsection{Holonomy groupoid $C^*$-algebra for a gauge and gravitational theory}\label{subsec Holgroupoid}

However none of the $^*$- or $C^*$-algebras of the previous sections contain a quantum analogue of the classical variable curvature. This is related to the degree of freedom \ref{listofoptions6}. In particular the classical Hamilton constraint, which is given by \eqref{eq Hamconstraint}, contains curvature. This Hamilton constraint cannot be quantised without changing this operator by some classical modifications until now. The aim of this project about \textit{Algebras of Quantum Variables in LQG} is to find a suitable algebra of quantum variables of the theory. This algebra is specified by the fact that the quantum Hamilton constraint operator is an element of (or affiliated with) the algebra, which is generated by certain holonomies along paths, quantum fluxes and the quantum analogue of curvature. The certain holonomies are given by generalised holonomy maps, which are a further development of the holonomy maps, which have been presented by Barrett \cite{Barrett91}. 

Barrett has presented a roadmap for the construction of the configuration space of Yang-Mills or gravitational theories. In this project these ideas will play a fundamental role. In general the quantisation of a gravitational theory in the context of LQG uses substantially the duality between infinitesimal objects like connections and curvature and integrated objects like holonomies or parallel transports. The ideas have been further developed by Mackenzie \cite{Mack05} in a more general context of Lie groupoids. In the context of LQG this duality will be reviewed briefly \cite[Section 2.2]{Kaminski5}, \cite[Section 3.2]{KaminskiPHD} by using the theory of Mackenzie. The next paragraphs give a short outline about these objects and how they can be used to construct a new algebra.

In this project a smooth connection is encoded in terms of a new holonomy map. This object is derived from a new object, which is called a \textit{path connection in a Lie groupoid}. The path connections have been studied originally by Mackenzie \cite{Mack05}. The concept of Mackenzie fits into the framework of holonomy mappings. The new holonomy map is called the \textit{general holonomy map in a Lie groupoid} and this map is, in particular, a continuous groupoid morphism from the path groupoid to a general Lie groupoid, which satisfies some new conditions.

Let $G$ be a Lie group, then $G$ over $\{e_G\}$ is a simple Lie groupoid. Furthermore, consider a certain path groupoid, which is called a \textit{path groupoid along tangent germs}. In \cite[ Section 3.3.4]{KaminskiPHD} it has been argued that the general holonomy map for a path groupoid along tangent germs in the groupoid $G$ over $\{e_G\}$ corresponds one-to-one to a path connection. Notice that the original holonomy map in \cite[Section 2.2]{Kaminski1}, \cite[Section 3.3.4]{KaminskiPHD} is defined by a groupoid morphism from the path groupoid to the groupoid $G$ over $\{e_G\}$, which satisfies no additional conditions. 

In particular a gauge theory will be studied in \cite[Section 2.1]{Kaminski5}, \cite[Section 3.1.4]{KaminskiPHD}. The \textit{generalised holonomy maps for a gauge theory} are certain continuous maps from a path groupoid to the gauge groupoid. The \textit{gauge groupoid} w.r.t. a principal fibre bundle $P(\Sigma,G,\pi)$ is indeed a particular Lie groupoid. These generalised holonomy maps correspond uniquely to a path connection, which is given as the integrated infinitesimal smooth connection over a lifted path in the gauge groupoid. Therefore in this context the generalised holonomy map for a gauge theory defines a parallel transport in a fixed principal fibre bundle. A fixed holonomy map for a gauge theory defines the \textit{holonomy groupoid for a gauge theory}. Notice that the original holonomy map is defined by a groupoid morphism from the path groupoid to the groupoid $G$ over $\{e_G\}$. Consequently, this holonomy map does not define a parallel transport in $P(\Sigma,G)$.  

The author of the project \textit{AQV} suggests to generalise the idea of Barrett. Then the set of all general holonomy maps along loops or paths in a Lie groupoid will be chosen as the configuration space of the theory. The new configuration space for example in the context of a pure gauge theory is given by the \textit{set of holonomy maps for a gauge theory} in \cite[Section 2.3.2]{Kaminski5}, \cite[Section 3.3.3]{KaminskiPHD}. In a more general context the \textit{set of holonomy maps along tangent germs for a arbitrary Lie groupoid} has been defined in \cite[Section 3.3.4]{KaminskiPHD}. 

A new $C^*$-algebra is given by the \textit{holonomy groupoid $C^*$-algebra for a gauge theory associated to a path connection} in \cite[Section 3]{Kaminski5},  \cite[Section 9.1]{KaminskiPHD}. In this framework the configuration space is given by the holonomy groupoid of a gauge theory. The algebra is defined in analogy to the group algebra of a locally compact group. The measure on this groupoid is inherited from the measure defined on the gauge groupoid. This is similar to the original approach in LQG, where the measure on the quantum configuration space is inherited from the Lie group $G$. 

The next step is to find a replacement of the curvature. The problem of implementing infinitesimal structures like infinitesimal diffeomorphisms and curvature arises from the special choice of the configuration variables. For the construction of the analytic holonomy $C^*$-algebra \cite[Chapter 3]{Kaminski1}, \cite[Chapter 6]{KaminskiPHD}, or the non-commutative analytic holonomy $C^*$-algebra \cite[Chapter 2]{Kaminski2}, \cite[Section 7.1]{KaminskiPHD} the original holonomy maps along paths are used. These maps are, in particular, not necessarily continuous groupoid morphisms from the path groupoid to the structure group $G$. In this usual approach infinitesimal objects like curvature cannot be treated as operators, which are contained in the algebra of quantum variables. In the new approach by using the theory of Mackenzie, the quantum curvature can be implemented as such an operator. 

There exists a generalised Ambrose-Singer theorem given by Mackenzie \cite{Mack05}, which states that the \textit{Lie algebroid of the holonomy groupoid} of a path connection is the smallest Lie algebroid, which is generated from the connections and the curvature. In \cite{Kaminski5}, \cite[Chapter 9]{KaminskiPHD} it will be used that there exists a left (or right) action of the exponentiated Lie algebroid elements on the holonomy groupoid $C^*$-algebra for a gauge theory. Hence, there is an action related to infinitesimal connections and curvature, since both objects are encoded as elements of this Lie algebroid. Hence, the quantum algebra of a gauge theory is generated by the $G$-valued quantum flux operators, the holonomy groupoid $C^*$-algebra for a gauge theory and the Lie algebroid of the holonomy groupoid for a path connection. The construction of this algebra is influenced by the cross-product construction, which will be studied in \cite{Kaminski2}\cite[Chapter 7]{KaminskiPHD}. 

There exists a cross-product algebra constructed from the left (or right) action of the Lie holonomy groupoid on the algebra $C(G)$, where $G$ denotes the structure group. The development in \cite[Section 3.2]{Kaminski5}, \cite[Section 9.2]{KaminskiPHD} will be based on Masuda \cite{MasudaI,MasudaII}. There the cross-product algebra will be shortly proposed. This new algebra contains the holonomy groupoid and $G$-valued quantum flux operator. The quantum curvature and the connections are not contained in this algebra. But the author of the project \textit{AQV} suggests that these operators are affiliated in a proper sense with the new algebra called the \textit{holonomy-flux groupoid $C^*$-algebra for a gauge theory}.

Summarising new basic quantum variables will be introduced in \cite{Kaminski5,KaminskiPHD}. One of the new quantum variables are the generalised holonomy maps. Consequently a new configuration space of the quantum theory of gravity is given by these maps. Finally, this modification allows a new development of algebras of quantum gravity, which are not comparable to the algebras presented in the previous sections. 

In a extension of the project  \textit{AQV} these constructions have to be generalised to a concrete principal fibre bundle of a gravitational theory. Furthermore, the new construction implies new problems, which can be studied in the future. Some of them are the following:
\begin{itemize}
 \item The new quantum analogue of the classical Hamilton constraint, which contains curvature, can be derived. In particular the theory of constraints and KMS-theory can be studied with respect to this new quantum Hamilton constraint.
 \item The algebras depend manifestly on the chosen principal fibre bundle. Consequently there is a problem of background independence of the theory (the gauge groupoid depends on the principal fibre bundle). The author of this project suggests to use the ideas of Brunetti, Fredenhagen and Verch \cite{BrunFredVerch01} to study this issue.
\end{itemize}


\section*{Acknowledgements}
The author gratefully acknowledges C. Fleischhack for checking spelling and grammar. Furthermore, thanks to C. Fleischhack, J. Brunnemann, J. Aastrup, G. Grimstrup, I. Androulidakis, T. Thiemann, J. Lewandowski and S. Woronowicz for discussions.  
The work has been supported by the Emmy-Noether-Programm (grant FL 622/1-1) of the Deutsche Forschungsgemeinschaft.

\end{document}